\documentclass[preprint,aps,amsmath,showpacs,showkeys,nofootinbib, 
superscriptaddress]{revtex4} 
\usepackage[dvips]{graphicx} 
 
\begin{document} 

\title{\textbf{$K^-$ nuclear potentials from in-medium chirally motivated 
models}} 

\author{A.~Ciepl\'{y}}
\email{cieply@ujf.cas.cz} 
\affiliation{Nuclear Physics Institute, 25068 \v{R}e\v{z}, Czech Republic} 

\author{E.~Friedman} 
\email{elifried@vms.huji.ac.il} 
\affiliation{Racah Institute of Physics, The Hebrew University, 
Jerusalem 91904, Israel} 

\author{A.~Gal} 
\email{avragal@vms.huji.ac.il} 
\affiliation{Racah Institute of Physics, The Hebrew University, 
Jerusalem 91904, Israel} 

\author{D.~Gazda} 
\email{gazda@ujf.cas.cz} 
\affiliation{Nuclear Physics Institute, 25068 \v{R}e\v{z}, Czech Republic} 

\author{J.~Mare\v{s}} 
\email{mares@ujf.cas.cz} 
\affiliation{Nuclear Physics Institute, 25068 \v{R}e\v{z}, Czech Republic} 

\date{\today} 

\begin{abstract} 

A self consistent scheme for constructing $K^-$ nuclear optical potentials 
from subthreshold in-medium ${\bar K}N$ $s$-wave scattering amplitudes 
is presented and applied to analysis of kaonic atoms data and to 
calculations of $K^-$ quasibound nuclear states. The amplitudes are 
taken from a chirally motivated meson-baryon coupled-channel model, 
both at the Tomozawa-Weinberg leading order and at the next to leading 
order. Typical kaonic atoms potentials are characterized by a real part 
$-{\rm Re}\:V_{K^-}^{\rm chiral}=85\pm 5$~MeV at nuclear matter density, 
in contrast to half this depth obtained in some derivations based on 
in-medium ${\bar K}N$ threshold amplitudes. The moderate agreement with 
data is much improved by adding complex $\rho$- and $\rho ^2$--dependent 
phenomenological terms, found to be dominated by $\rho ^2$ contributions 
that could represent ${\bar K}NN\to YN$ absorption and dispersion, 
outside the scope of meson-baryon chiral models. Depths of the real 
potentials are then near 180~MeV. The effects of $p$-wave interactions 
are studied and found secondary to those of the dominant $s$-wave 
contributions. The in-medium dynamics of the coupled-channel model is 
discussed and systematic studies of $K^-$ quasibound nuclear states are 
presented. 

\end{abstract} 



\maketitle 


\section{Introduction} 
\label{sec:intro} 

A key issue in studying in-medium $K^-$ meson interactions concerns 
the strength of the attractive $K^-$ nuclear potential \cite{npa804}. 
Related topical questions involve (i) the underlying free-space $\bar KN$ 
interaction and whether or not it can realistically support $K^-$ nuclear 
clusters (see Ref.~\cite{weise10} for a recent review), and (ii) the role of 
$K^-$ mesons in multistrange self-bound matter \cite{GFGM07} and in compact 
stars \cite{SSS08}. An order of magnitude estimate of the nuclear potential 
$V_{K^-}$ is provided by the leading-order (LO) Tomozawa-Weinberg (TW) vector 
term of the chiral effective meson-baryon Lagrangian \cite{TW66} which in the 
Born approximation gives  
\begin{equation} 
\label{eq:chiral} 
V_{K^-}=-\frac{3}{8f_{\pi}^2}~\rho\approx -57~\frac{\rho}{\rho_0} \:\:\:\:\:\: 
{\rm (in~MeV)}, 
\end{equation} 
where $\rho$ is the nuclear density, $\rho_0=0.17~{\rm fm}^{-3}$, 
and $f_{\pi} \approx 93$ MeV is the pion decay constant. 
This attraction is doubled, roughly, within chirally based coupled-channel 
$\bar K N$--$\pi\Sigma$--$\pi\Lambda$ calculations that produce dynamically 
a $\bar K N$ quasibound state loosely identified with the $\Lambda(1405)$ 
resonance \cite{weise96}. Deeper potentials, in the range ${\rm Re}\:V_{K^-}
(\rho_0)\sim -$(150--200) MeV are obtained in comprehensive global fits to 
$K^-$-atom strong-interaction shifts and widths by introducing empirical 
density dependent effective $K^-N$ amplitudes \cite{FGB93,FGMC99,MFG06,FG07}. 
Such strongly attractive potentials are expected to generate $K^-$ nuclear 
quasibound states which could prove relatively narrow once the strong 
transition ${\bar K} N\to\pi\Sigma$ becomes kinematically forbidden for 
binding energies exceeding about 100 MeV, as conjectured by Akaishi and 
Yamazaki \cite{AY02}. Experimentally, we mention the $K^-$ quasibound 
signals claimed for $K^-pp$ \cite{FINUDA05,DISTO10} at and below the 
$\pi\Sigma N$ threshold. However, these reported signals are quite broad, 
at variance with the underlying physics. In contrast to the indications of 
a deep $K^-$ potential, considerably shallower potentials, ${\rm Re}\:V_{K^-}
(\rho_0)\sim -$(40--60) MeV, are obtained for zero kinetic-energy kaons by 
introducing self energy (SE) contributions to the in-medium $K^- N$ threshold 
scattering amplitude, within a self-consistent procedure that includes in 
particular the potential $V_{K^-}$ thus generated \cite{RO00,CFGM01}. 

In a recent Letter \cite{CFGGM11} we reported on new, self consistent 
calculations of $K^-$ quasibound states that lead to deep $K^-$ nuclear 
potentials, considerably deeper than the `shallow' potentials deduced in 
Refs.~\cite{RO00,CFGM01}. 
The basic idea is to identify the $K^-N$ {\it subthreshold} energy domain 
required for the construction of $V_{K^-}$. For kaonic atoms, essentially 
at the $K^-$ nuclear threshold, this was explored during the 1970s by Wycech 
\cite{wycech71}, Bardeen and Torigoe \cite{BTo72} and Rook \cite{rook75} 
who noted the dominance of the subthreshold $\bar K N$ quasibound state 
$\Lambda(1405)$ in causing the in-medium $\bar K N$ scattering amplitude 
to become more attractive as one goes to subthreshold $K^-N$ energies. 
In our Letter \cite{CFGGM11} we applied this idea, introducing a new self 
consistency requirement, to a comprehensive study of kaonic atoms that uses 
scattering amplitudes derived from a chirally motivated coupled channel 
meson-baryon Lagrangian \cite{CS10}. Here we expand on these recent 
calculations to provide more details on derivation, systematics and results. 
In addition to the next to leading-order (NLO) model CS30 used in the Letter, 
in the present work we report on a new LO model TW1 fitted to the new 
SIDDHARTA values of shift and width of the $1s$ state in the $K^-$ hydrogen 
atom \cite{SIDD11}. 
The paper is organized as follows: in Sec.~\ref{sec:self} we describe a self 
consistent scheme of handling in-medium subthreshold $K^-N$ scattering 
amplitudes used in the construction of $V_{K^-}$. In Sec.~\ref{sec:model} we 
discuss the derivation of in-medium scattering amplitudes in both models TW1 
and CS30. Some details are relegated to an Appendix. In Sec.~\ref{sec:atoms} 
we discuss kaonic atom calculations, and in Sec.~\ref{sec:bound} we discuss 
calculations of $K^-$ nuclear quasibound states. Sec.~\ref{sec:concl} 
concludes the work with a brief summary of the main results.

\section{Handling $K^-N$ subthreshold amplitudes} 
\label{sec:self} 

In the single-nucleon approximation, the $K^-$ potential in nuclear matter 
of density $\rho$ is given in terms of the in-medium $K^-N$ scattering 
amplitude $F_{K^-N}$, 
\begin{equation} 
V_{K^-}=-\:\frac{2\pi}{\omega_K}\:(1+\frac{\omega_K}{m_N})\:
F_{K^-N}(\vec p,\sqrt{s};\rho)\:\rho, 
\label{eq:t} 
\end{equation} 
where $F_{K^-N}(\vec p,\sqrt{s};\rho\to 0)$ reduces to the free-space two-body 
$K^-N$ c.m. forward scattering amplitude $F_{K^-N}(\vec p,\sqrt{s})$ and the 
nucleon energy $E_N$ is approximated by its mass $m_N$ in the kinematical 
factor in front of $F_{K^-N}$. Here, $\vec p$ is the relative $K^-N$ momentum 
and $s=(E_K+E_N)^2-({\vec p}_K+{\vec p}_N)^2$ is the Lorentz invariant 
Mandelstam variable $s$ which reduces to the square of the total $K^-N$ 
energy in the two-body c.m. frame. In the laboratory frame, $E_K=\omega_K$. 
Before constructing $V_{K^-}$ for use in actual calculations, we need to 
prescribe how to interpret in Eq.~(\ref{eq:t}) the two-body arguments 
$\vec p$ and $\sqrt{s}$ of the in-medium scattering amplitude. For $s$-wave 
amplitudes, the momentum dependence arises through the magnitude $p$ of the 
relative momentum $\vec p$ which near threshold is approximated by 
\begin{equation} 
\vec p=\xi_N{\vec p}_K-\xi_K{\vec p}_N, \;\;\;\xi_{N(K)}=m_{N(K)}/(m_N+m_K). 
\label{eq:vecp} 
\end{equation} 
Averaging over angles, the square of $\vec p$ assumes the form 
\begin{equation} 
p^2 \rightarrow \xi_N\xi_K(2m_K\frac{p_N^2}{2m_N}+2m_N\frac{p_K^2}{2m_K}). 
\label{eq:p^2} 
\end{equation} 
For $\sqrt{s}$ we note that ${\vec p}_K+{\vec p}_N=0$ in the two-body c.m. 
system, but ${\vec p}_K+{\vec p}_N \neq 0$ in the nuclear laboratory system 
which nearly coincides with the $K^-$--nucleus c.m. system. Averaging over 
angles yields $({\vec p}_K+{\vec p}_N)^2 \to (p_K^2+p_N^2)$. Near threshold, 
neglecting quadratic terms in the binding energies $B_K=m_K-E_K$, 
$B_N=m_N-E_N$, we have 
\begin{equation} 
\sqrt{s} \approx E_{\rm th} - B_N - B_K - \xi_N\frac{p_N^2}{2m_N} - 
\xi_K\frac{p_K^2}{2m_K}, 
\label{eq:sqrts} 
\end{equation} 
where $E_{\rm th}=m_N+m_K$. To transform the momentum dependence into density 
dependence, the nucleon kinetic energy $p_N^2/(2m_N)$ is approximated in the 
Fermi gas model by $T_N(\rho/\rho_0)^{2/3}$, with $T_N=23.0$ MeV, and the 
$K^-$ kinetic energy $p_K^2/(2m_K)$ is identified in the local density 
approximation with $-B_K -{\rm Re}\:{\cal V}_{K^-}(\rho)$, where 
${\cal V}_{K^-}=V_{K^-}+V_c$ and $V_c$ is the $K^-$ finite-size Coulomb 
potential. Under these approximations, Eqs.~(\ref{eq:p^2}) and 
(\ref{eq:sqrts}) become 
\begin{equation} 
p^2 \approx \xi_N\xi_K[2m_K T_N(\rho/\rho_0)^{2/3}-
2m_N(B_K+{\rm Re}\:{\cal V}_{K^-}(\rho))], 
\label{eq:p^2final} 
\end{equation} 
where both terms on the r.h.s. are positive for attractive $V_{K^-}$, and 
\begin{equation} 
{\sqrt{s}} \approx E_{\rm th} - B_N - \xi_N B_K - 
15.1(\frac{\rho}{\rho_0})^{2/3}+\xi_K{\rm Re}\:{\cal V}_{K^-}(\rho)  
\label{eq:sfinal} 
\end{equation} 
(in MeV), where {\it all} the terms following $E_{\rm th}$ on the r.h.s. are 
negative, thus implementing the anticipated downward energy shift into the 
$K^-N$ subthreshold energy region. Eq.~(\ref{eq:sfinal}) is used in most of 
the bound state applications below as is, although we also checked the effect 
of implementing gauge invariance through the substitution $\sqrt{s} \to 
\sqrt{s}-V_c$. Gauge invariance often is not implemented in the solution 
of the free-space Lippmann-Schwinger equations of underlying chiral models 
simply because its effects on the two-body meson-baryon system are negligible. 

We note that the $K^-$ nuclear potential $V_{K^-}$ appears as an argument 
in expressions (\ref{eq:p^2final}) and (\ref{eq:sfinal}) for $p^2$ and 
$\sqrt{s}$, respectively, which in turn serve as arguments in expression 
(\ref{eq:t}) for this same $V_{K^-}$. This suggests to calculate $V_{K^-}$ 
self consistently within a scheme in which the downward energy shift into 
the $K^-N$ subthreshold energy region is density dependent and is controlled 
by the outcome self-consistent $V_{K^-}(\rho)$. In the corresponding sections 
below we elaborate on the self consistency scheme which is applied to the 
solution of the wave equation satisfied by in-medium $K^-$ mesons.

\section{In-medium $\bar K N$ amplitudes} 
\label{sec:model} 

The synergy of chiral perturbation theory and coupled channel $T$-matrix 
resummation techniques provides successful description of $\bar{K}N$ 
interactions at low energies \cite{weise10}. In our approach we employ 
chirally motivated coupled-channel $s$-wave potentials that are taken in 
a separable form, 
\begin{equation}  
V_{ij}(p,p';\sqrt{s})=\sqrt{\frac{1}{2\omega_i}\frac{M_i}{E_i}}\; g_{i}(p)
             \; \frac{C_{ij}(\sqrt{s})}{f_{\pi}^2} \;
             g_{j}(p') \sqrt{\frac{1}{2\omega_j}\frac{M_j}{E_j}} \;\; ,
 \;\;\;\;\; g_{j}(p)=\frac{1}{1+(p/ \alpha_{j})^2} \; ,
\label{eq:Vpot} 
\end{equation} 
with $E_i$, $M_i$ and $\omega_i$ denoting baryon energy, baryon mass 
and meson energy in the c.m. system of channel $i$. The coupling matrix 
$C_{ij}$ is determined by chiral SU(3) symmetry. The parameter $f_{\pi} 
\sim 100$ MeV represents the pseudoscalar-meson decay constant in the chiral 
limit, and the inverse range parameters $\alpha_{i}$ are fitted to the low 
energy $\bar{K}N$ data. The indices $i$ and $j$ run over the meson-baryon 
coupled channels $\pi\Lambda$, $\pi\Sigma$, $\bar{K}N$, $\eta\Lambda$, 
$\eta\Sigma$ and $K\Xi$, including all their appropriate charge states. 
Details of the free-space version of this model are given in Ref.~\cite{CS10}. 
Here we summarize its essential points with emphasis on in-medium 
modifications. 

The chiral symmetry of meson-baryon interactions is reflected in the structure 
of the $C_{ij}$ coefficients derived directly from the Lagrangian. The exact 
content of the matrix elements up to second order in the meson c.m. kinetic 
energies was specified already in Ref.~\cite{KSW95}. In practice, one often 
considers only the leading order TW interaction \cite{TW66} with energy 
dependence given by 
\begin{equation}  
C_{ij}(\sqrt{s}) = - C_{ij}^{\rm TW} (2\sqrt{s} -M_{i} -M_{j})/4. 
\label{eq:CTW} 
\end{equation} 
The structure constants $C_{ij}^{\rm TW}$ are listed in Ref.~\cite{OR98}. 
We note that this relativistic prescription differs from the one 
adopted in models derived from a chiral Lagrangian formulation for static 
baryons \cite{CS10,KSW95} and expanded strictly only to second order 
in meson energies and quark masses. There, the energy dependence form 
$(2\sqrt{s}-M_{i}-M_{j})$ is replaced by $(\omega'_{i}+\omega'_{j})$ where 
the primed meson energies $\omega'_{j}$ include a relativistic correction: 
$\omega'_{j}=\omega_{j}+(\omega_{j}^{2}-m^{2}_{j})/(2M_{0})$, with $m_{j}$ 
denoting the meson mass in channel $j$ and where $M_0$ is the baryon mass in 
the chiral limit. In principle, approaches based on different formulations of 
the chiral Lagrangian should give identical results for physical observables. 
However, this is true only when one sums up an infinite series of relevant 
Feynman diagrams to all orders in $q$, and need not hold at a given 
perturbative order. In other words, models based on different Lagrangian 
formulations, or models that differ from each other in prescribing how to 
treat terms beyond leading order, may give within reasonable limits different 
predictions for physical observables. 

The scattering amplitudes corresponding to the separable potentials 
(\ref{eq:Vpot}) are also of a separable form 
\begin{equation} 
F_{ij}(p,p';\sqrt{s})=g_{i}(p)f_{ij}(\sqrt{s})g_{j}(p'), 
\label{eq:sep} 
\end{equation} 
with the same form factors $g_{i}(p)$ and $g_{j}(p')$, and where the 
{\it reduced} scattering amplitude $f_{ij}$ is given explicitly by 
\begin{equation} 
f_{ij}(\sqrt{s})=-\frac{1}{4\pi f_{\pi}^{2}} \sqrt{\frac{M_{i}M_{j}}{s}}
\left[(1-C(\sqrt{s})\cdot G(\sqrt{s}))^{-1} \cdot C(\sqrt{s})\right]_{ij}\;. 
\label{eq:ampl} 
\end{equation} 
Here the meson-baryon propagator $G(\sqrt{s})$ is diagonal in the channel 
indices $i$ and $j$. When the elementary $\bar{K}N$ system is submerged in 
the nuclear medium one has to consider Pauli blocking and self energies 
(SE) generated by the interactions of mesons and baryons with the medium. 
Thus, the propagator $G(\sqrt{s})$ and the reduced amplitudes 
$f_{ij}(\sqrt{s})$ become dependent on the nuclear density $\rho$. 
The intermediate state Green's function is calculated as 
\begin{equation} 
G_{i}(\sqrt{s};\rho)=\frac{1}{f_{\pi}^2}\frac{M_i}{\sqrt{s}}\int_{\Omega_{i}
(\rho)}\frac{d^{3}{\vec p}}{(2\pi)^{3}}\frac{g_{i}^{2}(p)}{p_{i}^{2}-p^{2}
-\Pi_{i}(\omega_{i},E_{i},\vec{p};\rho)+{\rm i}0}\;\; . 
\label{eq:Green} 
\end{equation} 
Here $\vec{p_i}$ is the on-shell c.m. momentum in channel $i$ and the 
integration domain $\Omega_{i}(\rho)$ is limited by the Pauli principle in 
the $\bar{K}N$ channels. Included in the denominator of the Green's function 
(\ref{eq:Green}) is the sum $\Pi_{i}$ of meson and baryon self energies in 
channel $i$. In particular the kaon SE $\Pi_K=2\omega_K V_{K^-}$, which serves 
as input in Eq.~(\ref{eq:Green}) and therefore also in Eq.~(\ref{eq:ampl}) 
for the output reduced amplitude $f_{K^-N}$, requires by Eq.~(\ref{eq:t}) 
the knowledge of this same output $f_{K^-N}$. This calls for a {\it self 
consistent} solution of the in-medium reduced scattering amplitudes $f_{ij}
(\sqrt{s},\rho)$ as was first suggested by Lutz \cite{lutz98}. In the present 
calculation, following Ref.~\cite{CFGM01}, the baryon and pion self energies 
were approximated by momentum independent potentials $V=V_{0}\:\rho/\rho_{0}$ 
with real and imaginary parts of $V_{0}$ chosen consistently from mean-field 
potentials used in nuclear structure calculations and in scattering 
calculations, respectively. Specifically, we adopted 
$V_{0}^{\pi}=(30-\rm{i}10)$ MeV, $V_{0}^{\Lambda}=(-30-\rm{i}10)$ MeV, 
$V_{0}^{\Sigma}=(30-\rm{i}10)$ MeV and $V_{0}^{N}=(-60-\rm{i}10)$ MeV. 

The free parameters of the separable-interaction chiral models considered 
in Ref.~\cite{CS10} and in the present work were fitted to the available 
experimental data on low energy $\bar{K}N$ interactions, consisting of 
$K^- p$ low-energy cross sections for elastic scattering and reactions 
to the $\bar{K^0}n$, $\pi^{+}\Sigma^{-}$, $\pi^{-}\Sigma^{+}$, $\pi^0\Lambda$ 
and $\pi^{0}\Sigma^{0}$ channels (as listed in Ref.~\cite{KSW95}). 
In addition, the accurately determined $K^- p$ threshold branching ratios 
$\gamma$, $R_c$, $R_n$ \cite{Mar81} provide a rather strict test for any 
quantitative model. Another stringent test is provided by the recent SIDDHARTA 
measured values $\Delta E_{1s}$ and $\Gamma_{1s}$ of the $K^-$--hydrogen atom 
$1s$ level shift and width \cite{SIDD11}. 

\begin{table} 
\caption{$K^-p$ threshold observables calculated in several free-space 
LO coupled-channel chiral models. The $K^-$--hydrogen atom $1s$ shift 
$\Delta E_{1s}$ and width $\Gamma_{1s}$ (in eV) marked by asterisks were 
obtained from the calculated $K^{-}p$ scattering length by means of a modified 
Deser-Trueman relation \cite{MRR04} and are compared to the SIDDHARTA measured 
values \cite{SIDD11}. The $K^-p$ threshold branching ratios $\gamma$, $R_c$, 
$R_n$ are from Ref.~\cite{Mar81}. The last two columns list the calculated 
$I=0$ $S$-matrix pole positions $z_1,z_2$ (in MeV) on the [$-,+$] second 
Riemann sheet of the complex energy plane.} 
\begin{ruledtabular} 
\begin{tabular}{cccccccc} 
 & $\Delta E_{1s}$ & $\Gamma_{1s}$ & $\gamma$ & $R_c$ & $R_n$ & $z_1$ & $z_2$ 
\\  \hline 
 TW1                & 323       & 659       & 2.36  & 0.636 & 0.183 & 
(1371,$-$54) & (1433,$-$25)  \\ 
 JOORM \cite{JOO03} & 275$^{*}$ & 586$^{*}$ & 2.30  & 0.618 & 0.257 & 
(1389,$-$64) & (1427,$-$17)  \\
 HW \cite{HW08}    & 270$^{*}$ & 570$^{*}$ & 1.80  & 0.624 & 0.225 & 
(1400,$-$76) & (1428,$-$17)  \\[2mm] 
 exp.  & 283 & 541 & 2.36 & 0.664 & 0.189 & -- & -- \\ 
 error ($\pm$) & 42 & 111 & 0.04 & 0.011 & 0.015 & -- & -- \\
\end{tabular} 
\end{ruledtabular} 
\label{tab:models} 
\end{table} 

In the present work we focus on a separable-interaction LO chiral model 
marked TW1, constructed by fitting just two parameters to the data, 
$f_{\pi}=113\pm 2$ MeV for the PS meson decay constant and $\alpha=701\pm 20$ 
MeV for the common inverse range parameter, both within one's theoretical 
expectations. Some characteristics of the TW1 model in comparison to other LO 
models are listed in Table \ref{tab:models}. These LO models include only the 
leading TW interaction \cite{TW66}, with interchannel couplings given by 
Eq.~(\ref{eq:CTW}). Also listed in the table are the positions $z_1,z_2$ of 
the two $I=0$ $S$-matrix poles that reside on the second Riemann sheet [$-,+$] 
of the complex energy manifold, where the signs are those of the imaginary 
parts of the c.m. momenta in the $\pi\Sigma$ and $\bar{K}N$ channels, 
respectively. Their origin may be traced to poles in decoupled $I=0$ channels, 
a $\pi\Sigma$ resonance pole $z_1^{(0)}$ and a $\bar{K}N$ quasibound state 
pole $z_2^{(0)}$. The $\pi\Sigma$--$\bar{K}N$ interchannel coupling moves 
the poles away from their zero-coupling position, the precise full-coupling 
position exhibiting some model dependence. It is remarkable that all the LO 
TW models listed in the table are in close agreement on the position of the 
upper pole $z_2$. This agreement is spoiled when NLO corrections that 
require additional low energy constants to be fitted to the experimental data 
are included in the interchannel couplings. In contrast, the position 
of the lower pole $z_1$ exhibits model dependence already in TW models. 
Generally, it is located much further away from the real axis than the pole 
$z_2$. The pole $z_2$ is usually relegated to the subthreshold behavior of 
the $K^{-}p$ amplitude and to the $\Lambda(1405)$ resonance observed in the 
$\pi\Sigma$ mass spectrum in $\bar{K}N$ initiated reactions. Nuclear medium 
effects on the poles $z_1,z_2$ are discussed in the Appendix. 

In Fig.~\ref{fig:aKp1} we show the energy dependence of the reduced elastic 
scattering amplitudes $f_{K^-p}$ and $f_{K^-n}$ in model TW1 in free space 
and for two versions of in-medium modifications (marked `with' and `without' 
SE). Recall that $f_{K^-p}=\frac{1}{2}(f_{\bar KN}^{I=0}+f_{\bar KN}^{I=1})$ 
is affected by the subthreshold $I=0$ $\Lambda(1405)$ resonance, 
whereas $f_{K^-n}=f_{\bar KN}^{I=1}$ is not affected. Indeed, the free-space 
amplitudes, in dashed lines, exhibit a marked difference between $K^-p$ and 
$K^-n$, with the former amplitude showing a typical resonance structure. 
The pronounced peak in Im~$f_{K^-p}$ and the change of sign in Re~$f_{K^-p}$ 
point to the existence of a quasibound state generated by the $I=0$ $\bar KN$ 
interaction closely below the $K^{-}p$ threshold. In contrast, the pure $I=1$ 
$K^{-}n$ amplitude displays hardly any energy dependence besides a smooth and 
slow decrease of the imaginary part upon going to subthreshold energies where 
phase space cuts it down. The free-space $K^-n$ interaction is weakly 
attractive and its in-medium renormalization, given by the other curves on 
the right-hand panels, is rather weak and exhibits little density dependence, 
in clear distinction to the in-medium effect on the $K^-p$ amplitudes shown 
on the left-hand panels. Here, in-medium Pauli blocking moves the $K^{-}p$ 
free-space resonance structure to higher energies, as demonstrated by the 
dot-dashed lines (marked `without SE') in the left panels of the figure which 
correspond to nuclear matter density $\rho_0=0.17$ fm$^{-3}$. The TW1 results 
obtained here with Pauli blocking fully agree with those obtained long ago by 
Waas, Kaiser, Weise \cite{weise96} and which are recoverable upon switching on 
their parameter set in our chiral formulation. In contrast, a very different 
pattern was presented by Ramos and Oset \cite{RO00}, most likely due to their 
on-shell treatment of the intermediate state propagator and the inclusion of 
a "nucleon hole" term. 

\begin{figure}[htb] 
\centering 
\begin{tabular}{cp{1cm}c} 
\includegraphics[height=9cm,width=7.5cm]{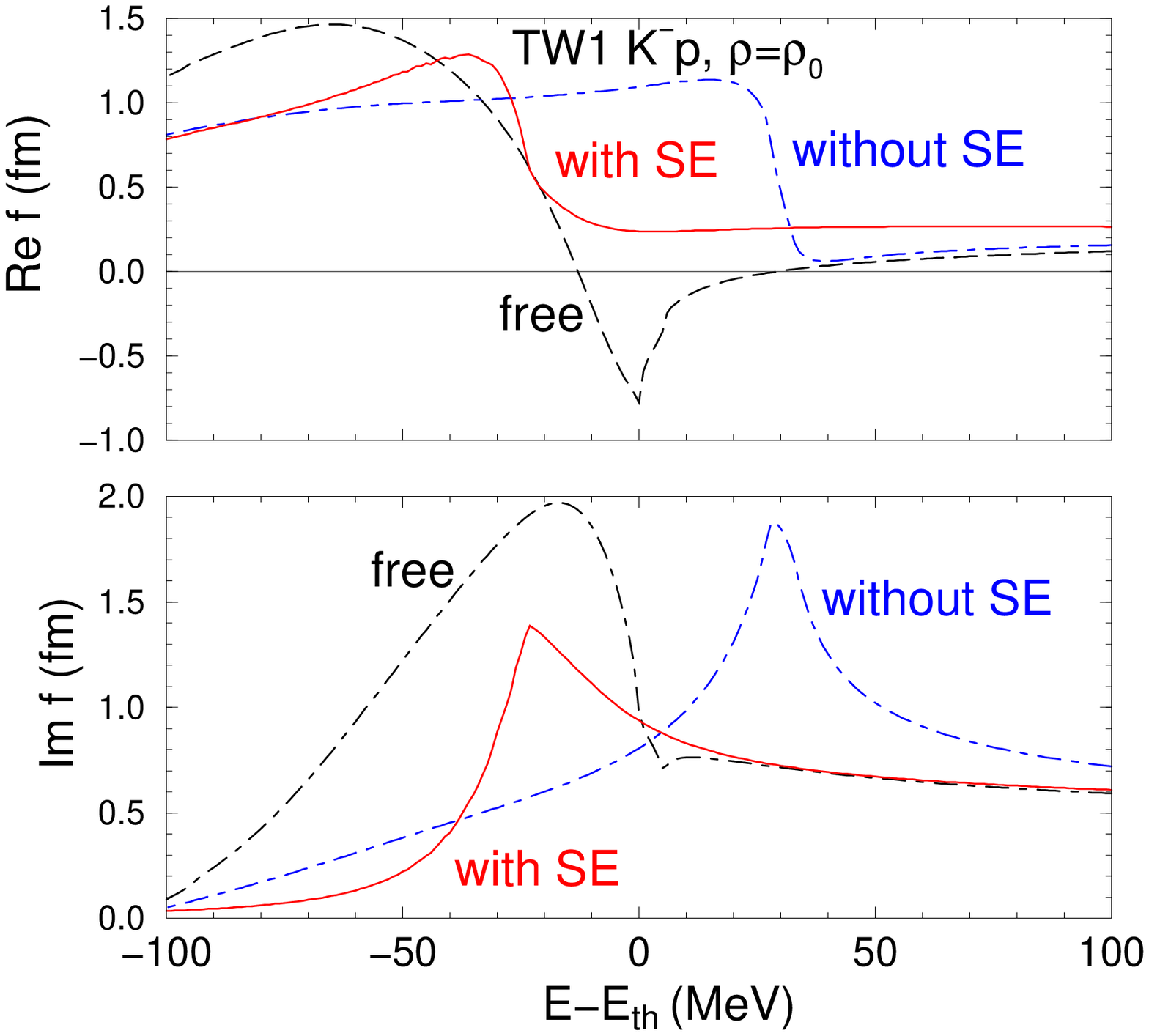} 
\hspace{1cm} 
\includegraphics[height=9cm,width=7.5cm]{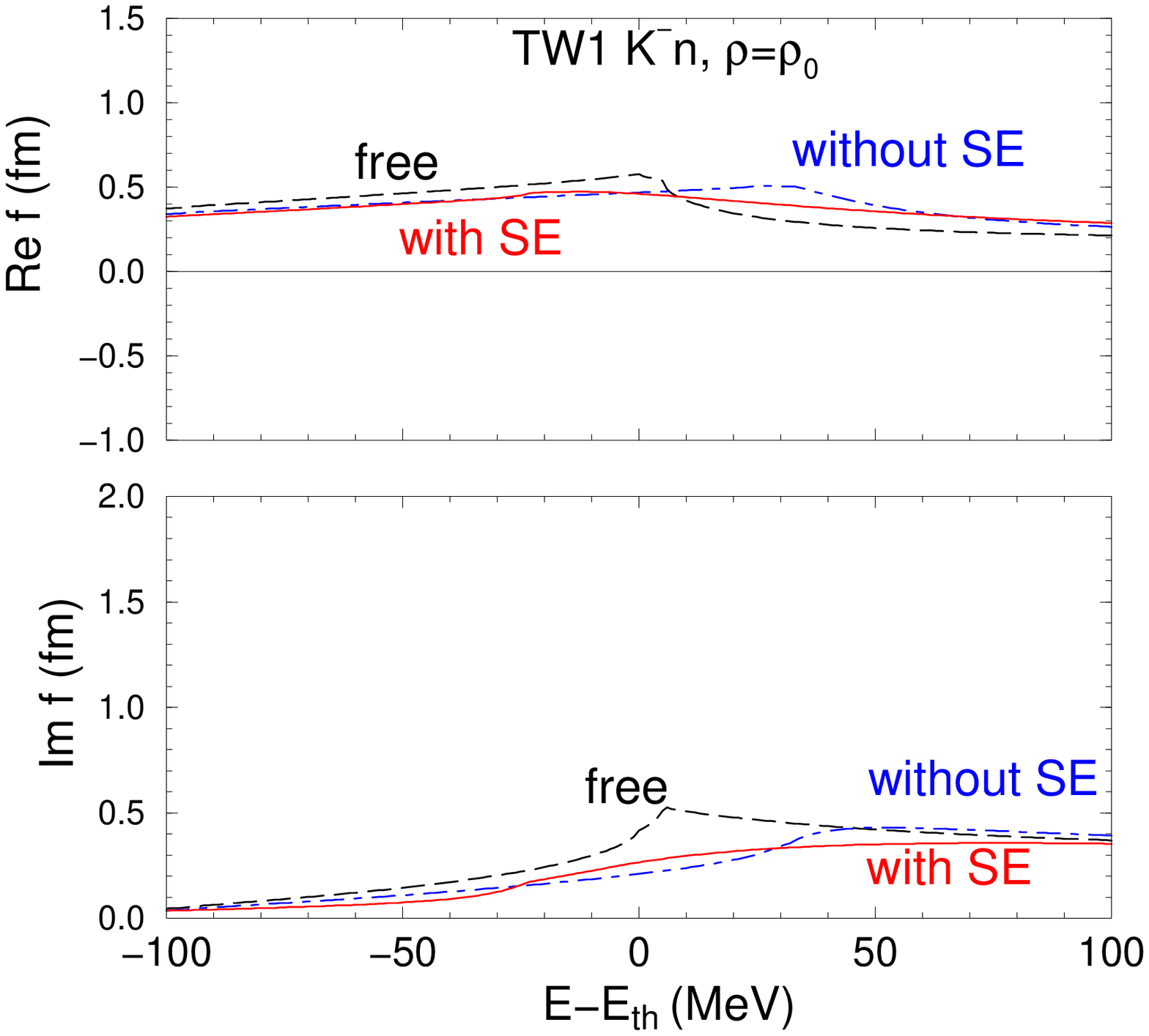} 
\end{tabular} 
\caption{(Color online) Energy dependence of the c.m. reduced amplitudes 
$f_{K^-p}$ (left panels) and $f_{K^-n}$ (right panels) in model TW1. The upper 
and lower panels refer to the real and imaginary parts of $f$, respectively. 
Dashed curves: free space, dot-dashed: Pauli blocked amplitude (without SE) 
at $\rho=\rho_0$, solid curves: including meson and baryon self energies 
(with SE) at $\rho=\rho_0$.} 
\label{fig:aKp1} 
\end{figure} 

The effect of combined Pauli blocking and hadron SE on the $K^-p$ amplitude 
is shown by the solid lines (marked `with SE') in Fig.~\ref{fig:aKp1}. 
The real part of the amplitude remains positive (attractive) in the whole 
energy range, in agreement with phenomenological analyses of kaonic 
atoms \cite{FGB93}, while the peak of the imaginary part moves back to 
approximately where it was in the free-space amplitude. The most striking 
feature of the model is the sharp increase in the real part of the amplitude 
when going to subthreshold energies, caused mainly by the introduction of 
kaon self energy in the propagator (\ref{eq:Green}) which is responsible 
for moving the resonant structure related to the $\Lambda(1405)$ back below 
the $\bar{K}N$ threshold. Consequently, the $K^{-}p$ interaction becomes 
much stronger at energies about 30 MeV below the $K^{-}p$ threshold with 
respect to its strength at threshold. This feature is missing in the 
in-medium calculations of Ref.~\cite{RO00} 
which get substantially different results than ours already when only Pauli 
blocking is accounted for. 

\begin{figure}[htb] 
\center{\includegraphics[width=0.55\textwidth]{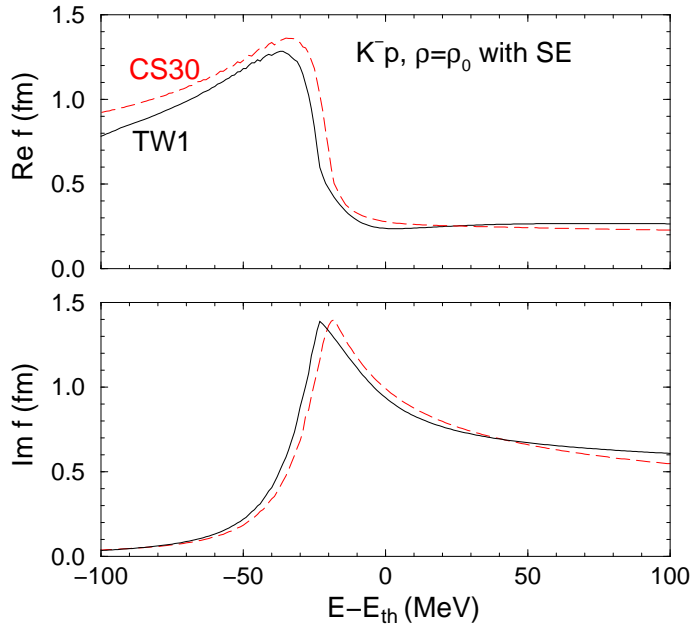}} 
\caption{(Color online) Energy dependence of the in-medium c.m. reduced 
amplitude $f_{K^-p}$ at nuclear matter density $\rho_0$ in models TW1 
(solid lines) and CS30 (dashed lines). The calculations include Pauli 
blocking and self energies.} 
\label{fig:aKp2} 
\end{figure} 

Although the simple LO TW1 model was used to demonstrate the nuclear medium 
effect on the $K^{-}p$ interaction in Fig.~\ref{fig:aKp1}, the same pattern is 
obtained within the NLO CS30 model of Ref.~\cite{CS10}. This is demonstrated 
in Fig.~\ref{fig:aKp2} where in-medium `with SE' $K^{-}p$ reduced scattering 
amplitudes generated in these two models are compared to each other at 
$\rho=\rho_0$. The differences between the two sets of curves are seen to be 
minute. 

\begin{figure}[thb] 
\begin{center} 
\includegraphics[width=0.6\textwidth]{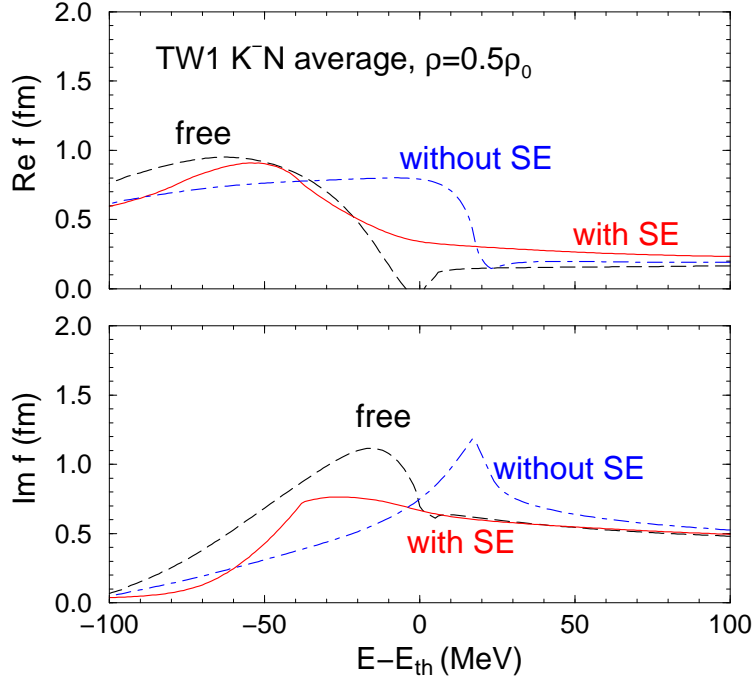} 
\caption{(Color online) Energy dependence of the c.m. reduced amplitude 
$f_{K^-N}$ (\ref{eq:isoscalar}) in model TW1 below and above threshold. 
Dashed curves: in free-space; dot-dashed curves: Pauli blocked amplitude 
at $0.5\rho_0$; solid curves: including meson and baryon self energies at 
$0.5\rho_0$.} 
\label{fig:TW1} 
\end{center} 
\end{figure}

To end this section we show in Fig.~\ref{fig:TW1}, for model TW1, the reduced 
scattering amplitude corresponding to the interaction of $K^-$ mesons with 
symmetric nuclear matter, 
\begin{equation} 
f_{K^-N}(\sqrt{s},\rho) = \frac{1}{2}[\:f_{K^-p}(\sqrt{s},\rho) + 
f_{K^-n}(\sqrt{s},\rho)\:], 
\label{eq:isoscalar} 
\end{equation} 
where $f_{K^-N}(\sqrt{s},\rho=0)\equiv f_{K^-N}(\sqrt{s})$. The free-space 
amplitude $f_{K^-N}(\sqrt{s})$, for $\rho=0$, is marked by dashed lines. 
Its imaginary part peaks about 15 MeV below the $\bar{K}N$ threshold, 
and its real part rapidly varies there from weak attraction above to strong 
attraction below threshold. While $f_{K^-N}(\sqrt{s})$ at and near threshold 
is constrained by data that serve to determine the parameters of the 
chiral model, the extrapolation to the subthreshold region may suffer 
from ambiguities depending on the applied model \cite{weise10}. 
Also shown in Fig.~\ref{fig:TW1} are two versions of in-medium reduced 
amplitudes $f_{K^-N}(\sqrt{s},\rho = 0.5\rho_0)$. One version, in dot-dashed 
lines (marked `without SE'), implements Pauli blocking in the intermediate 
$\bar KN$ states for $\rho\neq 0$. The resulting $f_{K^-N}$ exhibits 
a resonance-like behavior about 20 MeV above threshold, in agreement 
with Ref.~\cite{weise96}. The other in-medium version, in solid lines 
(marked `with SE'), adds self consistently meson and baryon self energies 
in intermediate states, as explained earlier. The resulting in-medium 
$f_{K^-N}$ is strongly energy dependent, with a resonance-like behavior about 
35 MeV below threshold. Similar results are obtained at full nuclear matter 
density $\rho_0=0.17$ fm$^{-3}$. We note that whereas the two in-medium 
reduced amplitudes shown in the figure are close to each other far below 
and far above threshold, they differ substantially at and near threshold. 
This applies also to the full amplitudes $F_{K^-N}$, Eq.~(\ref{eq:sep}), 
since the form factors $g(p)$ remain the same in the transition from 
free-space to in-medium separable amplitudes. At threshold, in particular, 
the real part of the `with SE' amplitude is about half of that `without SE', 
corresponding to a depth $-{\rm Re}\:V_{K^-}(\rho_0)\approx 40$~MeV, 
in rough agreement with Ref.~\cite{RO00}.

\section{$K^-$ atom calculations} 
\label{sec:atoms} 

Strong interaction level shifts and widths in kaonic atoms have been for 
decades a source of precise data on the $K^-$ nuclear interaction near 
threshold. Particularly instructive are so-called `global' analyses when 
data for many nuclei across the periodic table are being analyzed together, 
usually with the help of optical potentials which are related to the nuclear 
densities \cite{FG07}. This type of analyses could reveal characteristic 
features of the interaction which, in turn, reflect on the underlying $K^-N$ 
interaction in the medium, for example, its energy and density dependence. 
It was shown already in 1993 \cite{FGB93} that with density-dependent 
empirical amplitudes within a `$t\rho $' approach to the optical potential, 
very good fits to the data were possible. Depths of the real potential were 
close to 180~MeV whereas fixed-$t$ models achieved inferior fits and the 
resulting potentials were half as deep. Later predictions of in-medium chiral 
models {\it at threshold} \cite{RO00} presented depths of only 50~MeV for the 
real potential at full nuclear density. This wide span of values has been 
termed the `deep {\it vs.} shallow' controversy in kaonic atoms \cite{BFr07}. 
While attention has been focused on depths of the potentials, little attention 
was paid to the other empirical finding \cite{FGB93}, namely, that the best-fit 
real potentials were not only deep but also {\it `compressed'} relative to the 
corresponding nuclear densities, with r.m.s. radii smaller than the nuclear 
r.m.s. radii.  This feature means that the real part of the underlying 
$K^-$-bound nucleon interaction increases with density, and it is shown below 
to be in line with the density dependence of the chiral model in-medium 
amplitudes employed in the present work. 

\subsection{Wave equation} 
\label{sec:waveq} 

The choice of $K^-$ wave equation follows naturally from the in-medium 
dispersion relation 
\begin{equation} 
\omega_K^2-{{\vec p}_K}^{~2}-m_K^2-\Pi_K({\vec p}_K,\omega_K,\rho) = 0, 
\label{eq:KG} 
\end{equation} 
where $\Pi_K({\vec p}_K,\omega_K,\rho)=2({\rm Re}\:\omega_K)V_{K^-}$ is the 
self energy (SE) operator for a $K^-$ meson with momentum ${\vec p}_K$ and 
energy $\omega_K$ \cite{WH08}. The Klein-Gordon (KG) dispersion relation 
(\ref{eq:KG}) leads in hadronic atoms applications to a KG equation 
satisfied by the $K^-$ wavefunction \cite{FG07}: 
\begin{equation} 
[\:\nabla^2-2\mu({\cal B}_K+V_c)+(V_c+{\cal B}_K)^2+4\pi(1+\frac{A-1}{A}
\frac{\mu}{m_N})F_{K^-N}({\vec p},\sqrt{s};\rho)\:\rho\:]\:\psi=0. 
\label{eq:waveq} 
\end{equation} 
Here, $\mu$ is the $K^-$-nucleus reduced mass, ${\cal B}_K=B_K+{\rm i}
{\Gamma_K}/2$ is a complex binding energy, including a strong interaction 
width $\Gamma_K$, and $V_c$ is the $K^-$ Coulomb potential generated by 
the finite-size nuclear charge distribution, including vacuum-polarization 
terms. 

\subsection{$s$ waves} 
\label{sec:swave} 

The first application of the scheme presented in Sec.~\ref{sec:self} 
for handling $K^-N$ amplitudes below threshold was to global analyses of 
strong-interaction effects in kaonic atoms. The data base was the same as in 
Ref.~\cite{FGB93} with 65 data points for targets from $^7$Li to $^{238}$U. 
In solving the KG equation (\ref{eq:waveq}), the ${\vec p}_K$ momentum 
dependence of $F_{K^-N}$ was transformed into density and energy dependence 
according to Eq.~(\ref{eq:p^2final}). Furthermore, proton and neutron 
densities were handled separately, replacing $F_{K^-N}(\sqrt{s},\rho)\rho(r)$ 
by an effective amplitude 
\begin{equation} 
\label{eq:effA} 
{\cal F}_{K^-N}^{\rm eff}(\sqrt{s},\rho)\rho(r)= F_{K^-p} 
(\sqrt{s},\rho)\rho_p(r)+F_{K^-n}(\sqrt{s},\rho)\rho_n(r), 
\end{equation} 
with $\rho_p$ and $\rho_n$ normalized to $Z$ and $N$, respectively, and 
$Z+N=A$. Two-parameter Fermi distributions (2pF) were used for both densities, 
with $\rho_p$ obtained from the known charge distribution by unfolding the 
finite size of the charge of the proton. For $\rho_n$ averages of the `skin' 
and `halo' forms of Ref.~\cite{TJL01} were chosen with the difference 
between r.m.s. radii given by $r_n-r_p= (N-Z)/A-0.035$~fm. The reduced 
amplitudes $f_{K^-p}$ and $f_{K^-n}$ were evaluated at $\sqrt{s}$ given by 
Eq.~(\ref{eq:sfinal}), where the atomic binding energy $B_K$ was neglected 
with respect to $B_N\approx 8.5$ MeV. A similar approximation was made in 
Eq.~(\ref{eq:p^2final}) for $p^2$ when using the form factors $g(p)$ 
of Eq.~(\ref{eq:sep}). 
The $K^-$-nucleus potentials 
were calculated by requiring self consistency in solving Eq.~(\ref{eq:sfinal}) 
with respect to ${\rm Re}\:V_{K^-}$, i.e., the value of ${\rm Re}\:V_{K^-}
(\rho)$ in the expression for $\sqrt s$ and in the form factors $g$ had to 
agree with the resulting ${\rm Re}\:V_{K^-}(\rho)$. That was done at each 
radial point and for every target nucleus in the data base. 

\begin{figure}[htb] 
\begin{center} 
\includegraphics[height=7cm,width=8cm]{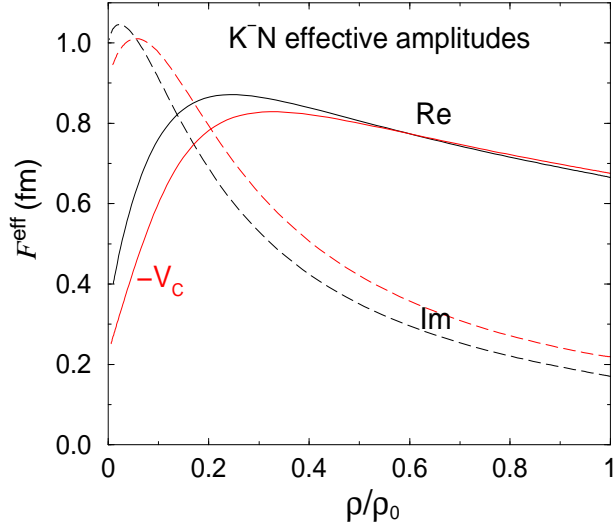} 
\caption{(Color online) Density dependence of the in-medium `without SE' CS30 
self consistent subthreshold amplitude ${\cal F}_{K^-N}^{\rm eff}$ for Ni.} 
\label{fig:effamp} 
\end{center} 
\end{figure} 

It is instructive to start by inspecting the effective amplitudes 
Eq.~(\ref{eq:effA}) obtained in the above self consistent procedure. 
Figure~\ref{fig:effamp} shows effective amplitudes for $K^-$ on Ni, 
calculated from the CS30 `without SE' $K^-N$ amplitudes, with and 
without the $\sqrt{s} \rightarrow \sqrt{s}-V_c$ substitution discussed 
in Sec.~\ref{sec:self}. The increase of ${\rm Re}\:{\cal F}_{K^-N}^{\rm eff}
(\rho)$ with density over the nuclear surface region combined with the 
decrease of ${\rm Im}\:{\cal F}_{K^-N}^{\rm eff}(\rho)$ are the underlying 
mechanisms behind the compression of the real part and inflation of the 
imaginary part of best-fit density-dependent phenomenological potentials 
\cite{FGB93}. Similar results for CS30 amplitudes that include SE were shown 
in Ref.~\cite{CFGGM11}. Although there are differences in details between 
the various models, the geometrical implications are robust. The decrease of 
${\rm Im}\:{\cal F}_{K^-N}^{\rm eff}$ with increasing density is unreasonably 
rapid, originating from the one-nucleon nature of the CS30 amplitudes, where, 
as seen in Fig.~\ref{fig:TW1} for the similar TW1 amplitude, the imaginary 
part practically vanishes around 80 MeV below threshold. We note that 
multi-nucleon absorption processes which become increasingly important 
at subthreshold energies are not included in the present approach. 
Since strong-interaction effects in kaonic atoms are dominated by the widths, 
the deficiency in the imaginary part of the amplitudes must be reflected when 
comparing predictions with experiment. This is indeed the case with $\chi ^2$ 
per point of about 10. 

\begin{figure}[t] 
\centering 
\begin{tabular}{cp{1cm}c} 
\includegraphics[height=9cm,width=7.5cm]{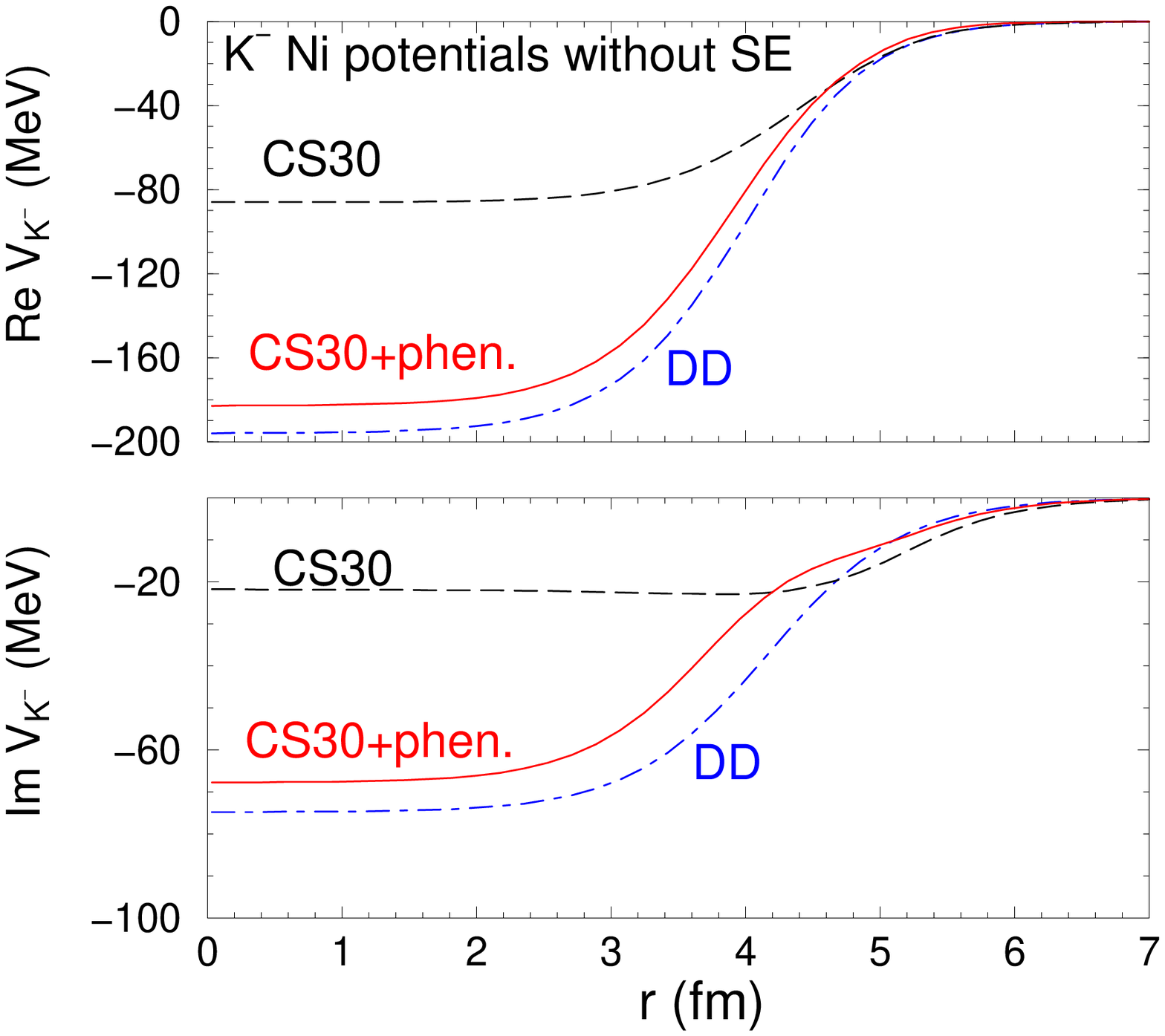} 
\hspace{1cm} 
\includegraphics[height=9cm,width=7.5cm]{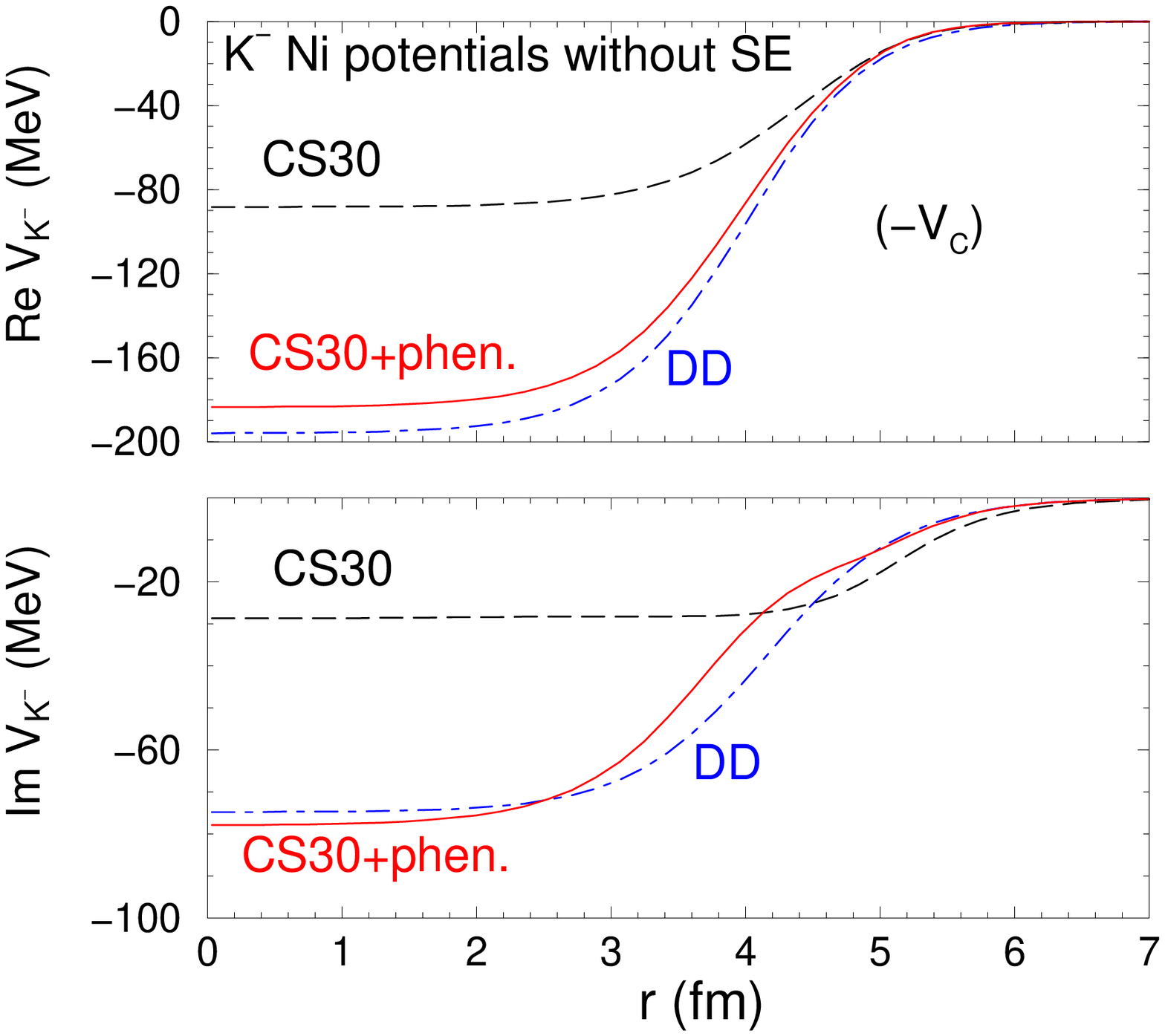} 
\end{tabular} 
\caption{(Color online) Left: $K^-$-nuclear potentials for $K^-$ atoms of Ni. 
Dashed curves: derived self-consistently from in-medium CS30 amplitudes; solid 
curves: plus phenomenological terms from global fits; dot-dashed curves: 
purely phenomenological DD potentials from global fits. Right: the same 
as on the left, but with the substitution $\sqrt{s}\rightarrow \sqrt{s}-V_c$ 
in ${\cal F}_{K^-N}^{\rm eff}(\sqrt{s},\rho)$.} 
\label{fig:potls} 
\end{figure} 

\begin{figure}[thb]  
\begin{center} 
\includegraphics[width=0.6\textwidth]{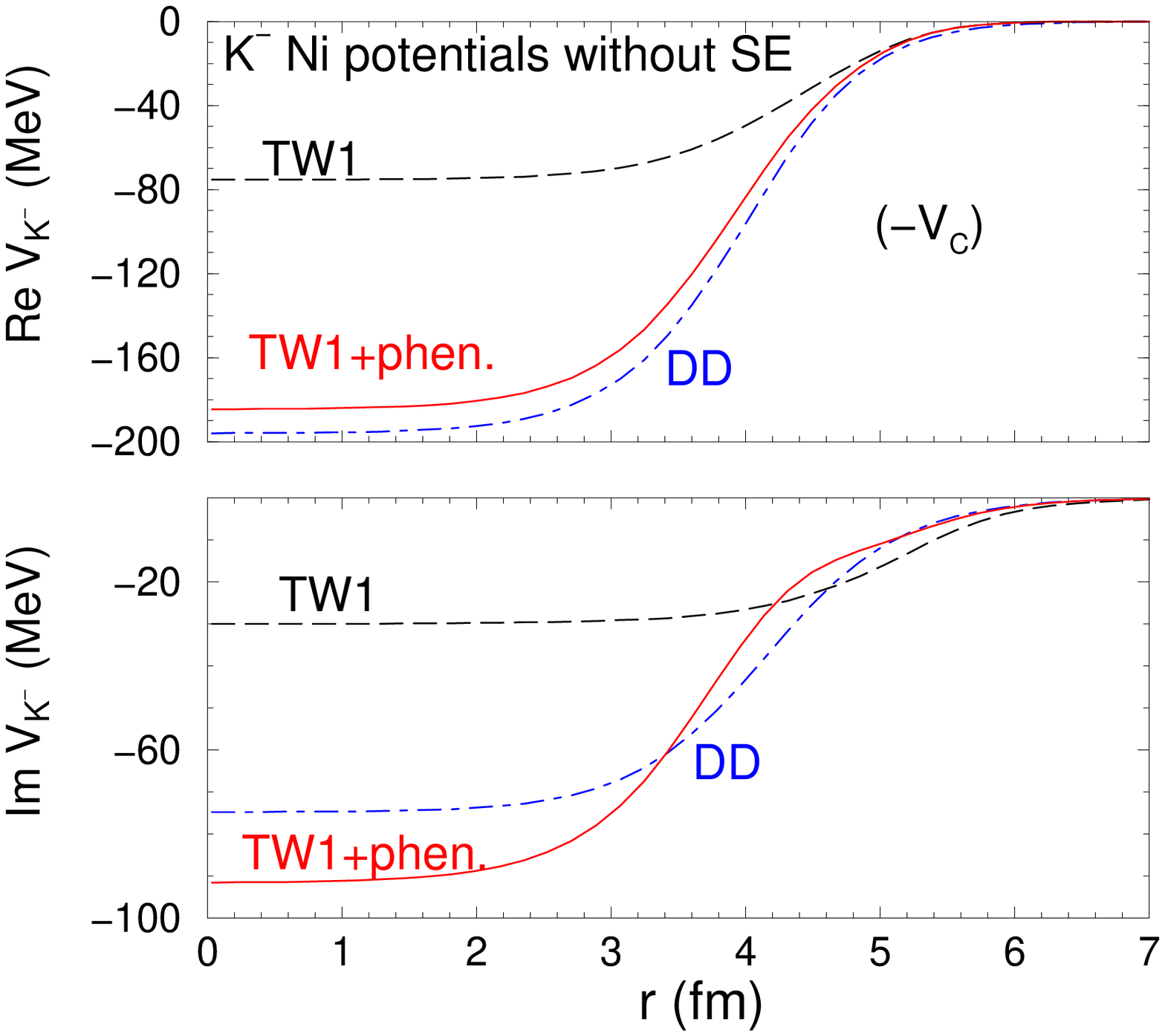} 
\caption{(Color online) $K^-$ nuclear real and imaginary potentials for $K^-$ 
atoms of Ni. Dashed curves: derived self-consistently from in-medium TW1 
amplitudes; solid curves: plus phenomenological terms from global fits; 
dot-dashed curves: purely phenomenological DD potentials from global fits.} 
\label{fig:atpotTW1} 
\end{center} 
\end{figure} 

Figure \ref{fig:potls} shows, as representative examples, several $K^-$--Ni 
potentials based on the CS30 `without SE' amplitudes, within the self 
consistent procedure described above. As a reference, the curves marked DD 
represent the best fit purely phenomenological density-dependent potentials 
\cite{FGB93} with $\chi ^2=103$ for 65 data points. The potential marked CS30 
is without any adjustable parameters and it differs substantially from the DD 
reference potential. Nevertheless its real part of $-$85 MeV is twice as deep 
as the shallow potential (not shown here) of Ref.~\cite{BGN00} which results 
from {\it threshold} values $f_{K^-N}(E_{\rm th},\rho)$, without going 
subthreshold. Figure~\ref{fig:potls} also demonstrates the effect of adding 
adjustable $\rho$ and $\rho^2$ terms to the CS30-based potentials, resulting 
in best-fit potentials $V_{K^-}^{\rm CS30+phen.}$ with $\chi ^2$ of around 
130--140 for 65 points, very close to $\chi ^2$ values achieved with 
a fixed-$t$ approach. It is seen that the resulting `CS30+phen.' potentials 
are close to the DD ones and we note that the additional terms, both real 
and imaginary, are dominated by  $\rho ^2$ terms which are required by the 
fit procedure and which are likely to represent ${\bar K}NN$ absorptive and 
dispersive contributions, respectively. Similar results hold for in-medium 
TW1 amplitudes, such as shown in Fig.~\ref{fig:TW1}. The resulting TW1 $K^-$ 
nuclear potentials are shown in Fig.~\ref{fig:atpotTW1}, exhibiting remarkable 
similarity to the CS30 $K^-$ nuclear potentials of Fig.~\ref{fig:potls}. 
We note that the addition of phenomenological terms lowers the resulting 
$\chi ^2$ to as low a value as 124 for 65 data points.

On first sight the additional `phen.' terms appear large in comparison with 
the one-nucleon based CS30 and TW1 potentials, particularly if one considers 
values of the potentials near the nuclear center. However, strong interaction 
effects in kaonic atoms are sensitive mostly to potential values near the 
nuclear surface \cite{BFr07}, about 3.5 to 5 fm in the examples shown. It is 
seen from the figures that over this range of radii the phenomenological part 
of the imaginary potential is of the order of 30\% of the starting values, 
consistent with the fraction of multi-nucleon absorptions estimated from 
experiments in emulsion and bubble chambers \cite{VVW77}. By the same token 
one may safely conclude that the data imply real potentials of depths 80--90 
MeV near the half-density radius. Finally a significant observation is that 
when the CS30 or the TW1 amplitudes are taken {\it at threshold}, then the 
additional phenomenological potential is no longer dominated by $\rho ^2$ 
terms. In particular, {\it negative} imaginary $\rho ^2$ terms are obtained, 
thus defying a two-nucleon absorption interpretation. 
The emerging phenomenology is similar to that for $V_{\pi^-}$ in pionic 
atom studies where theoretically motivated single-nucleon contributions 
are augmented by phenomenological $\rho^2$ terms representing $\pi NN$ 
processes \cite{EE66}. More work is required to justify microscopically 
the size of the $\rho^2$ kaonic atom contributions suggested by successful 
$V_{K^-}^{\rm chiral+phen.}$ potentials. 

\subsection{Adding $p$ waves} 
\label{sec:pwave} 

Next we turn to the question of whether kaonic atom data support contributions 
from a $p$-wave term in the $K^-N$ interaction and, for reference, we first 
checked the effect of including such a phenomenological term in a $t\rho $ 
potential. A $p$-wave term was added to the simplest $t\rho$ $s$-wave 
potential as follows \cite{FG07}: 
\begin{equation} \label{eq:EE1} 
2\mu V_{K^-N}(r) = q(r) + \vec \nabla \cdot \alpha(r) \vec \nabla 
\end{equation} 
with $q(r)$ its $s$-wave part given by 
\begin{equation}\label{eq:swave} 
q(r)  =  -4\pi(1+\frac{\mu}{m_N})b_0[\rho_n(r)+\rho_p(r)] 
\end{equation} 
and the $p$-wave part given by 
\begin{equation} 
\alpha (r)  =  4\pi(1+\frac{\mu}{m_N})^{-1} c_0[\rho_n(r)+\rho_p(r)]. 
\end{equation} 
Terms proportional to $\rho_n(r)-\rho_p(r)$ are neglected here. 

\begin{table}[thb] 
\caption{Results of $t\rho $ global fits to kaonic atoms data.} 
\begin{ruledtabular} 
\begin{tabular}{ccccc}  
~$\chi ^2$ (N=65)~ &~ Re $b_0$ (fm)~ &~ Im $b_0$ (fm)~ &~ Re $c_0$ (fm$^3$)~ & 
~Im $c_0$ (fm$^3$) \\ \hline 
 132 & 0.60 $\pm$ 0.05 & 0.94 $\pm$ 0.05 & --& -- \\ 
 110 & 0.73 $\pm$ 0.09 & 0.59 $\pm$ 0.09 & $-0.79\pm 0.32$ & 0.75 $\pm$ 0.30 \\ 
\end{tabular} 
\end{ruledtabular} 
\label{tab:strho}  
\end{table} 

It is seen from Table \ref{tab:strho} that an improved fit to the data is 
obtained with some of the absorption shifted from the $s$-wave term into 
the $p$-wave term which is repulsive and, thereby, the $s$-wave attraction 
required to fit the data is enhanced. However, this could also be just 
a numerical effect of the $\chi ^2$ fit process `compressing' the real part 
of the otherwise $s$-wave $t\rho$ potential, as noted above. 

The empirical $p$-wave term may be compared, for example, with the $K^-p$ 
$p$-wave amplitude of Weise and H\"{a}rtle \cite{WH08} which is dominated 
by the $I=1$ $\Sigma$(1385) subthreshold resonance. Over the energy range 
between about 1385 MeV and the $K^-N$ threshold at 1432 MeV the $K^-p$ 
$p$-wave amplitude is approximated there by 
\begin{equation} 
\label{eq:Kpwave} 
c_{K^-p}=\frac{\sqrt s ~\gamma_1}{s_0-s-{\rm i}\sqrt s~\Gamma(\sqrt s)}+d 
\end{equation} 
with $\sqrt s_0$=1385 MeV, $\gamma_1$=0.42/$m_K^3$, $\Gamma(\sqrt s)
\approx 40$ MeV and a background term $d$=0.06 fm$^3$. Considering that 
$c_{K^-n}=2c_{K^-p}$ for an $I=1$ dominated amplitude, then for 
$\rho_p\approx\rho_n$ and neglecting $\rho_n-\rho_p$ terms in the empirical 
potential, $c_{K^-p}$ is to be multiplied by 3/2 in order to compare with the 
above $c_0$. Table~\ref{tab:pwave} shows calculated values for a `microscopic' 
$c_0^m=(3/2)c_{K^-p}$ obtained for $\sqrt s$ given by Eq.~(\ref{eq:sfinal}), 
here applied to Ni. 

\begin{table}[thb] 
\caption{Values of a `microscopic' $p$-wave amplitude $c_0^m=(3/2)c_{K^-p}$ 
[see Eq.~(\ref{eq:Kpwave})] to be compared with the empirical $c_0$ of 
Table~\ref{tab:strho}.} 
\begin{ruledtabular} 
\begin{tabular}{cccc} 
~$\rho/\rho_0$~&~$\sqrt s$ (MeV)~&~Re $c_0^m$ (fm$^3$)~&~Im $c_0^m$ (fm$^3$)\\ 
\hline 
0 & 1432 & $-$0.09 & 0.08 \\ 
0.25 & 1420 & $-$0.12 & 0.12 \\ 
0.50 & 1404 & $-$0.16 & 0.25 \\ 
0.75 & 1392 & $-$0.06 & 0.44 \\ 
~1.00 & 1382 &  0.10   & 0.49 \\ 
\end{tabular} 
\end{ruledtabular} 
\label{tab:pwave} 
\end{table} 

It is seen that at threshold the empirical $c_0$ is an order of magnitude 
too large compared to its `microscopic' counterpart $c_0^m$. Averaging over 
subthreshold energies is unlikely to produce agreement between the two. 

The natural next step was to  include $c_{K^-p}$ of Eq.~(\ref{eq:Kpwave}) 
in the subthreshold evaluation of the $s$-wave potential to create also 
a $K^-N$-based $p$-wave potential. This was done at each radial point for 
the local density and the $\sqrt s$ obtained self-consistently for the 
dominant $s$-wave potential. Without any adjustable parameters it reduced 
the CS30-based $\chi ^2$ from $\approx$ 10 per point to about 6 per point. 
Including also two scaling factors, for the resonance part and for the 
background part of Eq.~(\ref{eq:Kpwave}) and searching on these parameters, 
yielded a scaling factor $-0.025\pm 0.029$ for the resonance and $3.5\pm 0.1$ 
for the background, with ${\chi}^2\approx 3$ per point. It means that within 
the subthreshold approach to the $K^-N$ interaction a resonance term in the 
$p$-wave interaction is not required to fit the data. 
This is consistent with the first comprehensive phenomenological analysis 
of $\bar KN$--$\pi Y$ coupled channels by Kim \cite{Kim67}, concluding that 
the $\Sigma(1385)$ is definitely not a $p$-wave $\bar KN$ bound state, but 
rather a $\pi\Lambda$ scattering resonance with very weak coupling to the 
$\bar KN$ channel. This conclusion was reinforced in a dispersion relation 
analysis by A.D. Martin \cite{Mar81} in which the $\bar KN$ channel coupled 
very weakly, compatible with zero coupling to the $\Sigma(1385)$ resonance. 
Finally, the $p$-wave amplitude $c_{K^-p}$ of Eq.~(\ref{eq:Kpwave}) was 
included in the 'CS30+phen.' fits, where $\rho$ and $\rho ^2$ terms were 
added to the CS30 potentials. Again the resonance term was found to vanish 
and only a small $p$-wave constant background term was acceptable. 
It is therefore concluded that fits to kaonic atom data do not require 
a resonant $p$-wave term within the subthreshold self consistent approach 
of the present work.

\section{Calculations of $K^-$ nuclear quasibound states} 
\label{sec:bound} 

Quasibound $K^-$ nuclear states in several nuclei across the periodic table 
were calculated in Refs.~\cite{MFG06,GFGM07} within the relativistic mean 
field (RMF) model for nucleons and antikaons. The energy independent 
$K^-$ nuclear real potential $V^{\rm RMF}_{K^-}$ was supplemented in 
these calculations by a phenomenological `$t\rho$' imaginary potential 
${\rm Im}\:V_{K^-}$ with energy dependence that accounted for the reduced 
phase space available for in-medium $K^-$ absorption. Two-nucleon absorption 
terms were also included. 

The present formulation differs fundamentally from these previous RMF 
calculations in that we use a $K^-$ nuclear potential $V_{K^-}$ given 
by Eq.~(\ref{eq:t}) in terms of energy and density dependent in-medium 
$K^-N$ scattering amplitudes $F_{K^-N}(\vec p,\sqrt{s},\rho)$ generated 
from a well defined coupled-channel chiral model. The momentum dependence 
of $F_{K^-N}$ was transformed into energy and density dependence 
using Eq.~(\ref{eq:p^2final}). The in-medium KG dispersion relation 
(\ref{eq:KG}) leads to a bound-state KG equation satisfied by the $K^-$ 
wavefunction which is written here in the form  
\begin{equation}
\left[ \nabla^2 + \omega_K^2 - m_K^2 + 4\pi \frac{\sqrt{s}}{m_N}
F_{K^-N}(\sqrt{s},\rho)\rho \right] \psi = 0, 
\label{eq:KGstrong} 
\end{equation} 
where 
\begin{equation} 
\omega_K = m_K - {\cal B}_K - V_c \; , 
\label{eq:omegaK} 
\end{equation}  
${\cal B}_K=B_K+{\rm i}\Gamma_K/2$ and $\sqrt{s}$ is given by 
Eq.~(\ref{eq:sfinal}) which now also includes the substitution 
$\sqrt{s} \to \sqrt{s}-V_c$: 
\begin{equation} 
{\sqrt{s}} \approx E_{\rm th} - B_N - \xi_N (B_K + V_c) 
- 15.1(\frac{\rho}{\rho_0})^{2/3} + \xi_K{\rm Re}\:V_{K^-}(\rho). 
\label{eq:sqrt} 
\end{equation} 
Equation (\ref{eq:KGstrong}) differs from the $K^-$ atom equation 
(\ref{eq:waveq}) by $A^{-1}$ correction terms. Since $B_K$ and 
$V_{K^-}(\rho)$ appear through Eq.~(\ref{eq:sqrt}) in the argument 
$\sqrt{s}$ of $F_{K^-N}$ (the latter is essentially $V_{K^-}$), 
it suggests a self consistency scheme in terms of {\it both} $B_K$ 
and $V_{K^-}(\rho)$ for solving the KG equation (\ref{eq:KGstrong}). 
In order to study the effect of energy and density dependencies of 
the argument $\sqrt{s}$ of the chiral $K^-N$ scattering amplitude 
$F_{K^-N}$, we first solved the KG equation in a static approximation, 
switching off the RMF self consistency cycle that accounts for the 
modification of the nuclear density by the strongly bound $K^-$ meson 
and its effect on the binding energy $B_K$. Self consistency with respect 
to $B_{K}$ and $V_{K^-}(\rho)$, however, remained operative in the static 
approximation. Realistic RMF density distributions $\rho(r)$ of the core 
nuclei were employed.
 
\begin{table}[htb] 
\begin{center} 
\caption{Binding energies $B_K$ and widths $\Gamma_K$ (in MeV) of $1s$ 
$K^-$ nuclear quasibound states in several nuclei, calculated using static 
RMF nuclear densities in Eq.~(\ref{eq:KGstrong}) and TW1 chiral amplitudes 
with (i) $\sqrt{s}=E_{\rm th}$ and (ii) $\sqrt{s}$ from Eq.~(\ref{eq:sqrt}), 
in both in-medium versions `no SE' and `+SE'. $K^-NN \rightarrow YN$ 
decay modes are excluded. Results of static RMF calculations of $B_K$, with 
a $K^-$ nuclear interaction mediated by vector mesons only, are shown for 
comparison in the last row.}
\begin{ruledtabular} 
\begin{tabular}{lc|ccccc} 
 & & $^{12}$C & $^{16}$O & $^{40}$Ca & $^{90}$Zr & $^{208}$Pb \\  \hline 
$E_{\rm th}$, no SE & $B_K$ &  61.1 & 57.5 & 83.4 & 96.0 & 104.8 \\ 
 &$\Gamma_K$ &  149.1 & 135.9 & 150.7 & 151.2 & 143.9 \\[2mm] 
$\sqrt{s}$, no SE  & $B_K$ & 40.9 & 42.4 & 58.5 & 69.5 & 77.6 \\ 
 &$\Gamma_K$ & 29.4 & 30.8 & 23.6 & 22.4 & 22.0 \\[2mm] 
$E_{\rm th}$, +SE & $B_K$ & (-0.9) & 6.4 & 25.0 & 39.0 & 53.4 \\ 
 &$\Gamma_K$ & (137.6) & 120.2 & 141.8 & 141.0 & 129.1\\[2mm] 
$\sqrt{s}$, +SE  &  $B_K$ & 42.4 & 44.9 & 58.8 & 68.9 & 76.3 \\ 
 &$\Gamma_K$ & 16.5 & 16.2 & 12.0 & 11.5 & 11.3 \\[2mm] 
 $V^{\rm RMF}_{K^-}$ &  $B_K$ & 49.1 & 47.7 & 60.5 & 69.6 & 76.8 \\ 
\end{tabular} 
\end{ruledtabular} 
\label{tab:bound1} 
\end{center}
\end{table}

In Table~\ref{tab:bound1}, we list binding energies $B_K$ and widths 
$\Gamma_K$ of $1s$ $K^-$ nuclear quasibound states obtained by solving 
Eq.~(\ref{eq:KGstrong}) self-consistently in several nuclei across the 
periodic table, using in-medium `no SE' and `+SE' TW1 subthreshold 
amplitudes with argument $\sqrt{s}$ given by Eq.~(\ref{eq:sqrt}) 
(denoted `$\sqrt{s}$' in the table). These values of $B_K$ and $\Gamma_K$ 
are compared to those calculated using threshold amplitudes without 
undergoing self consistency cycles (denoted `$E_{\rm th}$' in the table). 
The table illustrates the peculiar role of energy dependence of the 
${\bar K}N$ scattering amplitudes. In the `no SE' case, when the in-medium 
effects consist only of Pauli blocking, the self consistent calculations 
with subthreshold amplitudes yield lower $B_K$ values compared to those 
calculated using threshold amplitudes. In contrast, in the `+SE' case, 
when hadron in-medium self energies are included, the self consistent 
calculations with subthreshold amplitudes yield considerably higher $B_K$ 
values compared to those calculated using threshold amplitudes (in which 
case the $1s$ state in $^{12}$C is even unbound). It is worth noting that 
the self consistent calculations of $B_K$ give very similar results in the 
`+SE' version to those in the `no SE' version, as could be anticipated 
from the deep-subthreshold portion of the scattering amplitudes shown in 
Fig.~\ref{fig:aKp1}. These $B_K$ values are also remarkably close to those 
calculated within a static RMF approach for nucleons and antikaons, when 
the $K^-$ nucleus interaction is mediated exclusively by vector mesons 
with purely vector SU(3) F-type couplings, as shown in the last row of 
Table~\ref{tab:bound1}. 

\begin{figure}[htb] 
\center{\includegraphics[width=0.6\textwidth]{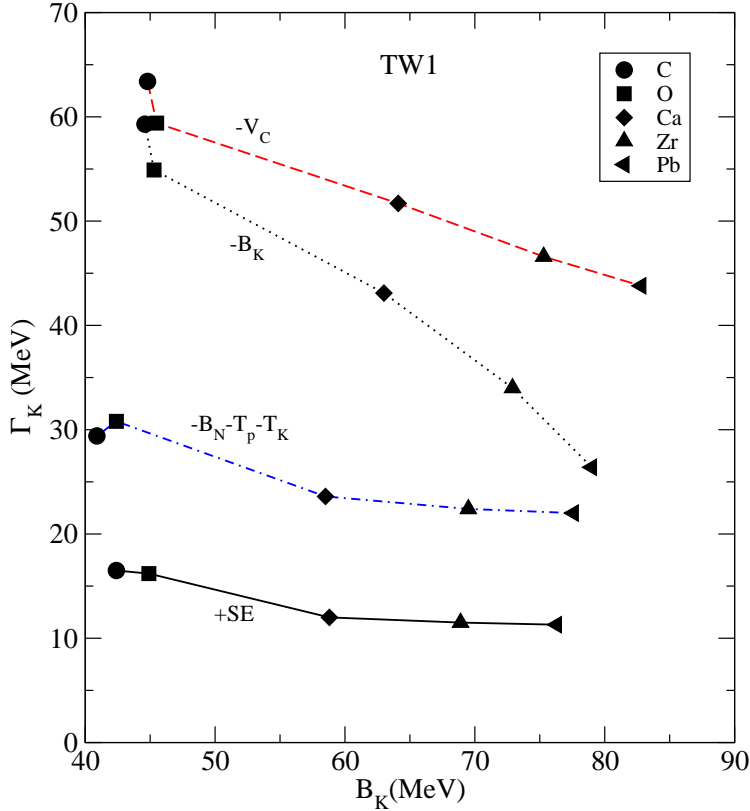}} 
\caption{(Color online) Binding energies $B_K$ and widths $\Gamma_K$ 
(in MeV) of $1s$ $K^-$ nuclear quasibound states, calculated by applying 
self-consistently several prescriptions of subthreshold $\sqrt{s}$ 
extrapolation with static RMF nuclear densities to in-medium TW1 
amplitudes (`no SE' unless specified `+SE'). $K^-NN \rightarrow YN$ 
decay modes are excluded.} 
\label{fig:bound1} 
\end{figure} 

The calculated widths displayed in Table~\ref{tab:bound1} represent only 
$K^-N \rightarrow \pi Y$ decays, accounted for by the coupled-channel chiral 
model. The widths are very large in both `no SE' and `+SE' in-medium versions 
when using threshold amplitudes, and are considerably smaller in the self 
consistent calculations using subthreshold amplitudes owing to the proximity 
of the $\pi\Sigma$ thresholds. In this case the `+SE' widths are about half 
of the `no SE' widths and approximately 10\% of those calculated using 
threshold amplitudes. 

\begin{figure}[htb] 
\center{\includegraphics[width=0.55\textwidth]{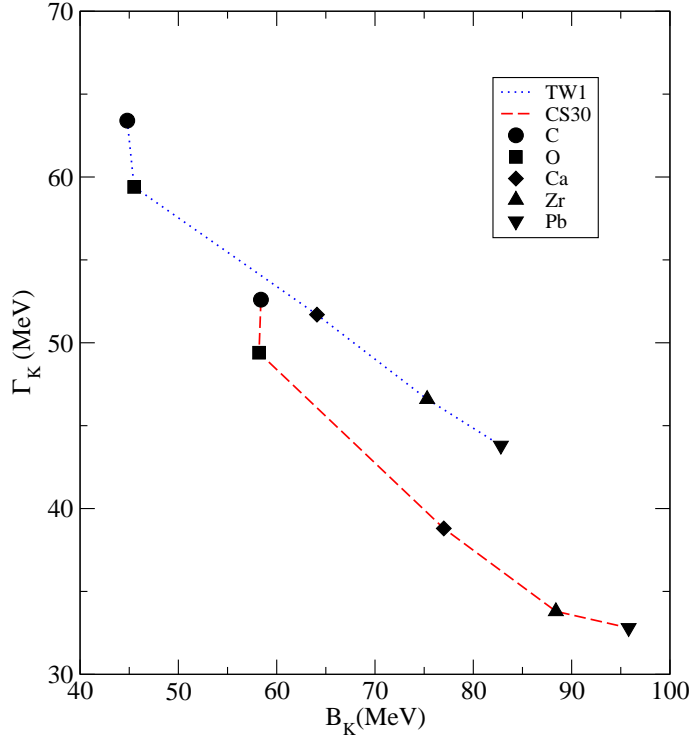}} 
\caption{(Color online) Binding energies $B_K$ and widths $\Gamma_K$ 
(in MeV) of $1s$ $K^-$ nuclear quasibound states, calculated by applying 
self-consistently the subthreshold extrapolation $\sqrt{s}=E_{\rm th}-B_K-V_c$ 
\cite{WH08} with static RMF nuclear densities to the TW1 and CS30 in-medium 
`no SE' amplitudes. $K^-NN \rightarrow YN$ decay modes are excluded.} 
\label{fig:bound2} 
\end{figure} 

The sensitivity of the calculated $K^-$ binding energies and widths to the 
specific form of the in-medium subthreshold extrapolation of $\sqrt{s}$ is 
demonstrated in Fig.~\ref{fig:bound1}. Here, $1s$ states in several nuclei 
are calculated self-consistently in the `no SE' version within the TW1 model 
for $\sqrt{s}=E_{\rm th}-B_K$ (dotted line), $\sqrt{s}=E_{\rm th}-B_K-V_c$ 
(dashed line), $\sqrt{s}$ of Eq.~(\ref{eq:sqrt}) (dot-dashed line), and for  
the latter $\sqrt{s}$ choice also in the `SE' version (full line). To lead the 
eye, each of the four lines connects ($B_K, \Gamma_K$) values in different 
core nuclei using one of the above forms for $\sqrt{s}$. It is seen that 
the specific form chosen to extrapolate $\sqrt{s}$ has a relatively small 
effect on the binding energies $B_K$, which vary within 5~MeV for a particular 
nucleus. In contrast, the widths are reduced significantly from about $55\pm 
10$~MeV to $14\pm 3$~MeV, when $\sqrt{s}$ is shifted further below threshold 
and the `SE' version which incorporates in-medium hadron self energies is 
applied. 

Figure~\ref{fig:bound2} illustrates the model dependence of $K^-$ nuclear 
quasibound state calculations by showing binding energies and widths of 
$1s$ states in several nuclei calculated self-consistently by applying the 
subthreshold extrapolation $\sqrt{s}=E_{\rm th}-B_K-V_c$ \cite{WH08} to the 
TW1 and CS30 in-medium `no SE' amplitudes. It is seen that the $K^-$ binding 
energies are more sensitive to the applied chiral model than to the form of 
subthreshold $\sqrt{s}$ extrapolation exhibited in Fig.~\ref{fig:bound1}. 
However, the difference in widths for a given nucleus is comparable to the 
differences due to the various forms of subthreshold $\sqrt{s}$ extrapolation 
shown there. The CS30 model produces higher binding energies and lower values 
of widths than in the TW1 model, with a difference of approximately 10~MeV. 
This systematics is explained by the stronger downward energy shift induced 
in CS30 with respect to TW1. 

\begin{table}[htb] 
\begin{center} 
\caption{Binding energies $B_K$ and widths $\Gamma_K$ (in MeV) of $1s$ $K^-$ 
nuclear quasibound states in several nuclei, calculated self-consistently 
using static RMF nuclear densities and the in-medium `no SE' version of TW1 
chiral amplitudes with (i) $\sqrt{s}=E_{\rm th}-B_K-V_c$ and (ii) $\sqrt{s}$ 
from Eq.~(\ref{eq:sqrt}), without and with $p$-wave amplitudes. 
$K^-NN \rightarrow YN$ decay modes are excluded.}
\begin{ruledtabular} 
\begin{tabular}{lc|ccccc} 
 & & $^{12}$C & $^{16}$O & $^{40}$Ca & $^{90}$Zr & $^{208}$Pb \\  \hline 
[WH] & $B_K$ & 44.8 & 45.5 & 64.1 & 75.3 & 82.8 \\ 
 &$\Gamma_K$ & 63.4 & 59.4 & 51.7 & 46.6 & 43.8 \\[2mm] 
 [WH], +$p$ wave & $B_K$ & 39.7 & 43.7 & 69.1 & 79.9 & 87.2 \\ 
 &$\Gamma_K$ & 85.6 & 73.6 & 55.5 & 46.7 & 44.3 \\[4mm] 
$\sqrt{s}$ & $B_K$ & 40.9 & 42.4 & 58.5 & 69.5 & 77.6 \\ 
 &$\Gamma_K$ & 29.4 & 30.8 & 23.6 & 22.4 & 22.0 \\[2mm] 
 $\sqrt{s}$, +$p$ wave & $B_K$ & 46.0 & 46.0 & 60.8 & 71.5 & 79.4 \\ 
 &$\Gamma_K$ & 27.5 & 29.6 & 22.4 & 21.3 & 21.0 \\ 
\end{tabular} 
\end{ruledtabular} 
\label{tab:bound2} 
\end{center} 
\end{table} 

Effects of including a $p$-wave ${\bar K}N$ interaction assigned 
to the $\Sigma(1385)$ subthreshold resonance are demonstrated in 
Table~\ref{tab:bound2} within the `no SE' in-medium version of TW1 
chiral amplitudes, for two subthreshold $\sqrt{s}$ extrapolations: 
[WH] denotes the form $\sqrt{s}=E_{\rm th}-B_K-V_c$ which was applied 
by Weise and H\"artle \cite{WH08} self-consistently to chiral ${\bar K}N$ 
amplitudes within a local density approximation to calculate $K^-$ nuclear 
$1s$ quasibound states in $^{16}$O and $^{208}$Pb; and $\sqrt{s}$ corresponds 
to the energy argument of Eq.~(\ref{eq:sqrt}). For the $p$-wave amplitude we 
adopted the parametrization (\ref{eq:Kpwave}) from Ref.~\cite{WH08}, used also 
in the previous section on kaonic atoms. The calculated binding energies and 
widths result from a delicate interplay between the energy dependent $s$-wave 
and $p$-wave amplitudes. The effect of $p$ waves is more pronounced in light 
nuclei where surface contributions are relatively more important, and it 
decreases with increasing size of the nucleus. The $p$-wave interaction leads 
to larger (smaller) widths in the [WH] ($\sqrt{s}$) version and increases the 
$K^-$ binding energies, with the exception of $^{12}$C and $^{16}$O in the 
[WH] subthreshold extrapolation where the substantially increased absorption 
acts repulsively to reduce $B_K$. 

\begin{table}[htb] 
\begin{center} 
\caption{Binding energies $B_K$ and widths $\Gamma_K$ (in MeV) of $1s$ $K^-$ 
nuclear quasibound states in several nuclei, calculated self-consistently 
using in-medium CS30 chiral amplitudes in the `no SE' version 
(first two sequences) and in the `+SE' version (last three sequences) 
for various subthreshold $\sqrt{s}$ extrapolations marked as in 
Table~\ref{tab:bound2}. A combined nucleons+antikaon RMF scheme is applied 
dynamically in the last two sequences, and $K^-NN \rightarrow YN$ decay modes 
are included in the last sequence (`+2N abs.').} 
\begin{ruledtabular} 
\begin{tabular}{lc|ccccc} 
 & & $^{12}$C & $^{16}$O & $^{40}$Ca & $^{90}$Zr & $^{208}$Pb \\  \hline
[WH] & $B_K$ & 58.4 & 58.2 & 77.0 & 86.7 & 95.8 \\
 &$\Gamma_K$ & 52.6 & 49.8 & 33.8 & 33.8 & 32.8 \\[2mm]
$\sqrt{s}$ & $B_K$ & 52.0 & 53.0 & 69.7 & 81.5 & 89.6 \\
 &$\Gamma_K$ & 19.6 & 21.6 & 14.4 & 13.6 & 14.0 \\[2mm]
 +SE  &  $B_K$ & 50.7 & 52.5 & 68.2 & 79.3 & 86.6 \\
 &$\Gamma_K$ & 13.0 & 12.8 & 10.9 & 11.0 & 10.9 \\[2mm]
 +dyn. &  $B_K$ & 55.7 & 56.0 & 70.2 & 80.5 & 87.0 \\
 &$\Gamma_K$ & 12.3 & 12.1 & 10.8 & 10.9 & 10.8\\[2mm]
 +2N abs. &  $B_K$ & 54.0 & 55.1 & 67.6 & 79.6 & 86.3 \\
  &$\Gamma_K$ & 44.9 & 53.3 & 65.3 & 48.7 & 47.3 \\ 
\end{tabular} 
\end{ruledtabular} 
\label{tab:bound3} 
\end{center} 
\end{table} 

Table~\ref{tab:bound3} presents binding energies and widths of $1s$ 
$K^-$ nuclear quasibound states in several nuclei across the periodic table, 
calculated self-consistently within in-medium versions of CS30 chiral 
amplitudes. The first two sequences denoted [WH] and $\sqrt{s}$ illustrate 
the role of subthreshold $\sqrt{s}$ extrapolation which affects particularly 
the widths $\Gamma_K$, in a similar pattern to that already shown for the TW1 
model in Table~\ref{tab:bound2}. The next three sequences exhibit the effects 
of successively sophisticating the calculations: first, the in-medium `no SE' 
amplitudes are dressed by self energies (denoted `+SE'), bringing the 
calculated widths further down (and marginally so the binding energies); then, 
the calculations are made dynamical (denoted `+dyn.') taking into account 
the polarization of the nuclear core by the strongly bound $K^-$, which 
produces higher binding energies $B_K$ and smaller widths $\Gamma_K$; and 
last, energy dependent imaginary $\rho^2$ terms are added self consistently 
to simulate two-nucleon $K^-NN \rightarrow YN$ absorption modes (denoted 
`+2N abs.') and their available phase space \cite{MFG06,GFGM07}. Whereas the 
binding energies decrease insignificantly, the resulting widths of order 
$\Gamma_K \sim 50$~MeV become comparable in light nuclei to the binding 
energies $B_K$.

\section{Conclusion} 
\label{sec:concl} 

In this work we have used several versions of in-medium $\bar K N$ scattering 
amplitudes constructed in a chirally motivated coupled channel separable 
potential model to derive self-consistently the $K^-$ nuclear potential 
for several bound state applications. The $\bar K N$ scattering amplitudes 
exhibit, invariably, a strong energy and density dependence below threshold, 
which reflects the dominant effect of the $\Lambda(1405)$ subthreshold 
resonance. This is precisely the energy region relevant for the self 
consistent construction of $V_{K^-}$ for kaonic atoms and for $K^-$ nuclear 
quasibound state calculations. It was found that kaonic atoms probe $\bar K N$ 
c.m. energies typically 30--50 MeV below threshold whereas $K^-$ nuclear $1s$ 
quasibound states reach considerably lower $\bar K N$ subthreshold energies. 
Thus, the chiral model versions used in the present work produced potential 
depths in the range $-{\rm Re}\:V_{K^-}^{\rm chiral}(\rho_0)\sim$~80--90 MeV 
in kaonic atoms, and somewhat deeper potentials of depths 100--110~MeV for 
$K^-$ nuclear quasibound states. By comparing the size and shape of our 
subthreshold $\bar K N$ scattering amplitudes with those of other chiral 
models, as discussed for example in Ref.~\cite{HW08}, we expect these results 
to hold generally in any coupled-channel chiral model constrained by low 
energy $K^-p$ data once our self consistency construction is applied. 
The density dependence of the resulting kaonic atom potentials is such that by 
adding adjustable phenomenological terms to be determined by fits to the data, 
the real part of the potential becomes twice as deep and the imaginary part 
about three times as deep due to a $\rho^2$--dominated complex term which 
could represent $\bar KNN\to YN$ dispersive and absorptive modifications. 
These substantial modifications at full nuclear density represent 
extrapolations from the nuclear surface region to which kaonic atoms 
are mostly sensitive and where such modifications appear more modest. 
More work is needed to explain the origin and test the existence of the 
sizable $\rho^2$ term. Finally, the effects of a $p$-wave interaction 
generated by the $\Sigma(1385)$ subthreshold resonance are found secondary 
to the effects of the $s$-wave interaction which is dominated by the 
$\Lambda(1405)$ subthreshold resonance.

\section*{Acknowledgments} 
Stimulating discussions with Wolfram Weise are gratefully acknowledged. 
This work was supported by the GACR Grant No. 202/09/1441, as well as by the 
EU initiative FP7, HadronPhysics2, under Project No. 227431.

\section*{Appendix: In-medium pole trajectories in model TW1} 

\begin{figure}[htb] 
\center{\includegraphics[width=0.6\textwidth]{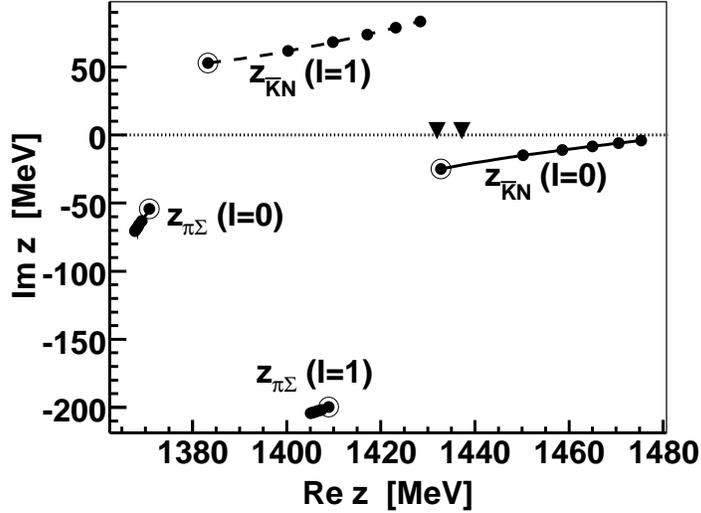}} 
\caption{Pole movements on the complex energy manifold due to the increased 
effect of Pauli blocking in model TW1. $I=0$ pole trajectories are marked by 
full lines, $I=1$ pole trajectories by dashed lines. Pole positions in free 
space are encircled and the bullets mark pole positions for $\rho=x\rho_0$ 
for increments of $x$ between 0 to 1, see text for more details. The solid 
triangles denote the $K^-p$ and ${\overline K}^0 n$ thresholds.} 
\label{fig:poles1} 
\end{figure}

The observed properties of in-medium $\bar{K}N$ interaction may be related to 
the dynamics of the $\Lambda(1405)$ resonance in the nuclear medium. This is 
demonstrated for model TW1 in Fig.~\ref{fig:poles1} which shows the motion in 
the complex energy plane of poles related to the $\pi\Sigma$ and $\bar{K}N$ 
channels upon increasing the nuclear density, including Pauli blocking but 
disregarding self energy insertions. The lower half of the energy plane, below 
the real axis, corresponds to the [$-,+$] Riemann sheet standardly referred to 
as the second Riemann sheet and accessed from the physical region by crossing 
the real energy axis in between the $\pi \Sigma$ and $\bar{K}N$ thresholds. 
The upper half of the energy plane (above the real axis) shows the [$+,-$] 
Riemann sheet, the third Riemann sheet, which does not allow for $\bar{K}N$ 
quasibound interpretation of poles located therein. The pole trajectories 
shown in the figure were calculated from the free-space pole positions 
(encircled dots) up to the pole positions at full nuclear density $\rho_{0}$. 
In addition to the two $I=0$ poles listed and discussed in 
Table~\ref{tab:models} of the main text, each of the decoupled $\pi\Sigma$ and 
$\bar{K}N$ channels also exhibits an $I=1$ pole, the one related to $\bar{K}N$ 
developing into a resonance residing on the [$+,-$] Riemann sheet and another 
one related to a $\pi\Sigma$ state residing on the [$-,+$] Riemann sheet. 
While the $I=1$ pole related to $\pi\Sigma$ lies too far from the real energy 
axis to affect any physical observable, the one related to $\bar{K}N$ is 
responsible for the peak structure in the real part of the $K^-n$ amplitude 
near threshold. However, the Riemann sheet location of this pole denies it 
of any quasibound interpretation. The $I=1$ poles persist also in the more 
involved NLO chiral models discussed in Ref.~\cite{CS10}. 

As expected, the nuclear medium has no significant impact on the 
position of poles related to the $\pi\Sigma$ channel. On the other hand, both 
poles related to the $\bar{K}N$ channel move to considerably higher energies 
as the density increases. The $I=0$ $\bar{K}N$ pole that affects most the 
$\bar{K}N$ scattering amplitude moves as high as about 1475 MeV, almost 
reaching the real energy axis. Since the pole is relatively far from the 
physical region due to the $\bar{K}N$ branch cut, the scattering amplitude 
exhibits a cusp instead of a proper resonance structure (see the dot-dashed 
line in Fig.~\ref{fig:aKp1} of the main text). When kaon self energy is 
implemented the pole moves back below the $\bar{K}N$ threshold, residing 
now in the [$+,-$] Riemann sheet. Since it remains relatively far from the 
physical region, one again gets a cusp structure as exhibited by the solid 
line in Fig.~\ref{fig:aKp1}.


\begin{thebibliography}{99} 

\bibitem{npa804} Contributions to {\it Recent advances in strangeness nuclear 
physics}, Eds. A.~Gal and R.S.~Hayano, Nucl. Phys. A \textbf{804} (2008) 
pp. 171-348. 

\bibitem{weise10} W.~Weise, Nucl. Phys. A \textbf{835}, 51 (2010), and 
references therein.  

\bibitem{GFGM07} D.~Gazda, E.~Friedman, A.~Gal, and J.~Mare\v{s}, 
Phys. Rev. C \textbf{76}, 055204 (2007); \textbf{77}, 045206 (2008); 
\textbf{80}, 035205 (2009). 

\bibitem{SSS08} J.~Schaffner-Bielich, S.~Schramm, and H.~St\"{o}cker, in 
{\it Proc. Int'l. School of Physics ``Enrico Fermi", Course \textbf{CLXVII}}, 
Eds. M.~Anselmino, {\it et al.} (IOS Press, Amsterdam, 2008) pp. 119--144; 
F.~\"{O}zel, G.~Baym, and T.~G\"{u}ver, Phys. Rev. D \textbf{82}, 101301(R) 
(2010). 

\bibitem{TW66} Y.~Tomozawa, Nuovo Cimento A \textbf{46}, 707 (1966); 
S.~Weinberg, Phys. Rev. Lett. \textbf{17}, 616 (1966). 

\bibitem{weise96} T.~Waas, N.~Kaiser, and W.~Weise, 
Phys. Lett. B \textbf{365}, 12 (1996); \textbf{379}, 34 (1996). 

\bibitem{FGB93} E.~Friedman, A.~Gal, and C.J.~Batty, Phys. Lett. B 
\textbf{308}, 6 (1993); Nucl. Phys. A \textbf{579}, 518 (1994). 

\bibitem{FGMC99} E.~Friedman, A.~Gal, J.~Mare\v{s}, and A.~Ciepl\'{y}, 
Phys. Rev. C \textbf{60}, 024314 (1999). 

\bibitem{MFG06} J.~Mare\v{s}, E.~Friedman, and A.~Gal, Nucl. Phys. A 
\textbf{770}, 84 (2006). 

\bibitem{FG07} E.~Friedman and A.~Gal, Phys. Rep. \textbf{452}, 89 (2007). 

\bibitem{AY02} Y.~Akaishi and T.~Yamazaki, Phys. Rev. C \textbf{65}, 044005 
(2002); T.~Yamazaki and Y.~Akaishi, Phys. Lett. B \textbf{535}, 70 (2002). 

\bibitem{FINUDA05} M.~Agnello {\it et al.} (FINUDA Collaboration), 
Phys. Rev. Lett. \textbf{94}, 212303 (2005). 

\bibitem{DISTO10} T.~Yamazaki {\it et al.} (DISTO Experiment), 
Phys. Rev. Lett. \textbf{104}, 132502 (2010). 

\bibitem{RO00} A.~Ramos and E.~Oset, Nucl. Phys. A \textbf{671}, 481 (2000). 

\bibitem{CFGM01} A.~Ciepl\'{y}, E.~Friedman, A.~Gal, and J.~Mare\v{s}, 
Nucl. Phys. A \textbf{696}, 173 (2001). 

\bibitem{CFGGM11} A.~Ciepl\'{y}, E.~Friedman, A.~Gal, D.~Gazda, and 
J.~Mare\v{s}, Phys. Lett. B \textbf{702}, 402 (2011). 

\bibitem{wycech71} S.~Wycech, Nucl. Phys. B \textbf{28}, 541 (1971). 

\bibitem{BTo72} W.A.~Bardeen and E.W.~Torigoe, Phys. Lett. B \textbf{38}, 135 
(1972). 

\bibitem{rook75} J.R.~Rook, Nucl. Phys. A \textbf{249}, 466 (1975). 

\bibitem{CS10} A.~Ciepl\'{y} and J.~Smejkal, Eur. Phys. J. A \textbf{43}, 
191 (2010). 

\bibitem{SIDD11} M.~Bazzi {\it et al.} (SIDDHARTA Collaboration), Phys. Lett. 
B \textbf{704}, 113 (2011), and references therein to previous experiments. 

\bibitem{KSW95} N.~Kaiser, P.B.~Siegel, and W.~Weise, Nucl. Phys. A 
\textbf{594}, 325 (1995). 

\bibitem{OR98} E.~Oset and A.~Ramos, Nucl. Phys. A \textbf{635}, 99 (1998).  

\bibitem{lutz98} M.~Lutz, Phys. Lett. B \textbf{426}, 12 (1998). 

\bibitem{Mar81} A.D.~Martin, Nucl. Phys. B \textbf{179}, 33 (1981), 
and references therein.

\bibitem{MRR04} U.-G.~Meissner, U.~Raha, and A.~Rusetsky, Eur. Phys. J. C 
\textbf{35}, 349 (2004).

\bibitem{JOO03} D.~Jido, J.A.~Oller, E.~Oset, A.~Ramos, and U.-G.~Mei{\ss}ner, 
Nucl. Phys. A \textbf{725}, 181 (2003). 

\bibitem{HW08} T.~Hyodo and W.~Weise, Phys. Rev. C \textbf{77}, 035204 (2008). 




\bibitem{BFr07} N.~Barnea and E.~Friedman, Phys. Rev. C \textbf{75}, 
022202(R) (2007). 

\bibitem{WH08} W.~Weise and R.~H\"{a}rtle, Nucl. Phys. A \textbf{804}, 
173 (2008). 

\bibitem{TJL01} A. Trzci\'{n}ska, J. Jastrz\c{e}bski, P. Lubi\'{n}ski,
F.J. Hartmann, R. Schmidt, T. von Egidy, and B. K{\l}os, 
Phys. Rev. Lett. \textbf{87}, 082501 (2001). 

\bibitem{BGN00} A.~Baca, C.~Garc\'{\i}a-Recio, and J.~Nieves, Nucl. Phys. A 
\textbf{673}, 335 (2000). 

\bibitem{VVW77} C.~Vander Velde-Wilquet, J.~Sacton, J.H.~Wickens, D.N.~Tovee, 
and D.H.~Davis, Nuovo Cimento \textbf{39} A, 538 (1977), and references 
therein. 

\bibitem{EE66} M.~Ericson and T.E.O.~Ericson, Ann. Phys. (NY) \textbf{36}, 
323 (1966). For a recent review see Ref.~\cite{FG07}. 

\bibitem{Kim67} J.K.~Kim, Phys. Rev. Lett. \textbf{19}, 1074 (1967). 




\end{thebibliography}
\end{document}